\documentclass[journal]{vgtc}                
\ifpdf
  \pdfoutput=1\relax                   
  \usepackage{graphicx}                
  \DeclareGraphicsExtensions{.pdf,.png,.jpg,.jpeg} 
\else
  \usepackage{graphicx}                
  \DeclareGraphicsExtensions{.eps}     
\fi%

\graphicspath{{figures/}{pictures/}{images/}{./}} 

\usepackage{microtype}                 
\PassOptionsToPackage{warn}{textcomp}  
\usepackage{textcomp}                  
\usepackage{mathptmx}                  
\usepackage{times}                     
\usepackage{cite}                      
\usepackage{tabu}                      
\usepackage{booktabs}                  
\usepackage{paralist}
\usepackage{stfloats}
\newcommand\score[1]{\includegraphics[height=0.9em]{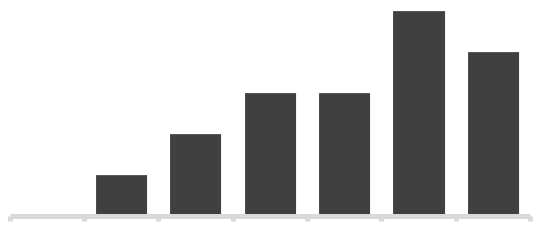}}
\newcommand\num[1]{108}
\usepackage{enumitem}
\usepackage[hyphens]{url}
\usepackage[hidelinks]{hyperref}
\usepackage{wrapfig}
\usepackage{setspace}

\onlineid{1232}

\vgtccategory{Research}
\vgtcpapertype{Empirical Study}

\title{What Makes a Data-GIF Understandable?}

\author{Xinhuan Shu, Aoyu Wu, Junxiu Tang, Benjamin Bach, Yingcai Wu, and Huamin Qu}
\authorfooter{
\item Xinhuan Shu is with the State Key Lab of CAD\&CG, Zhejiang University
and Hong Kong University of Science and Technology. A part of this
work was done when Xinhuan Shu was a visiting student supervised by
Yingcai Wu at Zhejiang University. E-mail:
xinhuan.shu@connect.ust.hk.
\item Aoyu Wu, and Huamin Qu are with the Hong Kong University of Science and Technology. E-mail: \{awuac, huamin\}@cse.ust.hk.
\item
 Junxiu Tang and Yingcai Wu are with the State Key Lab of CAD\&CG, Zhejiang University. Yingcai
Wu is the corresponding author. E-mail: \{tangjunxiu, ycwu\}@zju.edu.cn.
\item
Benjamin Bach is with Edinburgh University. E-mail: bbach@inf.ed.ac.uk.
}

\shortauthortitle{Biv \MakeLowercase{\textit{et al.}}: Global Illumination for Fun and Profit}

\abstract{GIFs are enjoying increasing popularity on social media as a format for data-driven storytelling with visualization; simple visual messages are embedded in short animations that usually last less than 15 seconds and are played in automatic repetition. In this paper, we ask the question, \textit{``What makes a data-GIF understandable?''} While other storytelling formats such as data videos, infographics, or data comics are relatively well studied, we have little knowledge about the design factors and principles for ``data-GIFs''. To close this gap, we provide results from semi-structured interviews and an online study with a total of 118 participants investigating the impact of design decisions on the understandability of data-GIFs. The study and our consequent analysis are informed by a systematic review and structured design space of 108 data-GIFs that we found online. Our results show the impact of design dimensions from our design space such as \hl{animation encoding}, context preservation, or repetition on viewers’ understanding of the GIF's core message. The paper concludes with a list of suggestions for creating more effective Data-GIFs.
} 


\CCScatlist{ 
 \CCScat{K.6.1}{Management of Computing and Information Systems}%
{Project and People Management}{Life Cycle};
 \CCScat{K.7.m}{The Computing Profession}{Miscellaneous}{Ethics}
}

\newcommand{\hl}[1]{\textcolor{black}{#1}}

\vgtcinsertpkg

\begin{document}



\maketitle
\section{Introduction}
The popularity of data-driven storytelling has grown rapidly these years. 
Media outlets such as The New York Times \cite{nyt}, The Washington Post \cite{washington}, FlowingData \cite{flowingdata}, and The Pudding \cite{thepudding}, are actively crafting data stories with visualizations. 
A large palette of data stories result in diverse genres in narrative visualization, such as infographics, comics, slideshows, and videos \cite{Segel2010}. 
Each genre possesses unique characteristics and affords opportunities for various communication scenarios, \textit{e.g.}, integrating text and graphics, leveraging linear and non-linear sequences, as well as combining both visual and auditory stimuli. 

In parallel with the advances of these established genres, we see a recent surge of attention and interests in ``data-GIFs'' \cite{DataGIFs, jsvine}, an emerging format that tells data stories with visualization. 
\hl{Typically, GIFs are short animations, usually less than 15 seconds, played in automatic repetition, and focusing on a single core message. 
These small animated visualizations are saved in the form of Graphics Interchange Format (GIF), making them easily accessible online.}
Data-GIFs are endowed with unique properties for communicating data insights. 
\hl{For example, compared to data videos~\cite{Amini2015}---a narrative visualization genre with longer playing time, more information, and potentially more complex narratives structures---data-GIFs are simpler with a shorter specific message and are more concise in terms of size and duration. 
They can be quickly loaded and automatically played and repeated, thereby supporting prompt reading and understanding. 
Moreover, GIFs capture viewers'  attention with motion effects~\cite{Bakhshi2016}.
Given the growing use and desirable properties, we argue data-GIFs are a distinct and promising genre for data-driven storytelling, worth of research and discussion.} 

However, the visualization research community has not yet given much consideration to data-GIFs. 
First, we lack an understanding of the current practices surrounding data-GIF designs. 
\textit{What factors should be considered when creating data-GIFs? }
Prior studies~\cite{Bakhshi2016, Jiang2018} have examined the content of animated GIFs, but their findings cannot fully contextualize data visualizations with different visual manifestations and communication goals. 
Given its roles for data-driven storytelling, the GIF design should involve specific considerations to craft data stories, compared to the known design principles for animated visualizations~\cite{Tversky2002, Heer2007}.
More importantly, it remains unclear \textit{what makes a data-GIF understandable to its intended audience}. 
We have little knowledge about the performance and effectiveness of data-GIFs for communicating data stories. 
For example, \textit{what does a data-GIF show over time?} and \textit{are animations easy to understand?}
Also, animation is commonly questioned to be inadequate to preserve the context and track the changes~\cite{Robertson2008}. 
\textit{How do data-GIFs present the frame sequence and facilitate the comprehension? }
In addition, as Munzner~\cite{munzner2015} claimed \textit{``giving people the ability to pause and replay the animation is much better than only seeing it a single time straight through''}, the performance of data-GIFs raises questions, since they do not allow the manipulation of the playing progress but automatically repeat the animation. 


In this paper, we set out to address the question: \textit{``What makes a data-GIF understandable?''}
Our work is the first to systematically explore data-GIFs as a distinct and promising medium for data-driven storytelling. 
To this end, we build a collection of \num{} real-world data-GIFs from a wide range of online websites such as social media, news portals, and personal blogs.
We then summarize the design practices based on the curated data-GIFs, whereby extracting the design factors from intra-frame and inter-frame perspectives, \textit{i.e.}, \textit{visualization types} and \textit{navigation progress}, \hl{\textit{animation encoding}}, \textit{context preservation}, and \textit{repetition}, respectively \hl{(Sec. \ref{sec:survey})}. 
The analysis of the design space helps us figure out specific characteristics that distinguish data-GIFs from other storytelling mediums.
\hl{We conduct a qualitative study through interviews (Sec. \ref{sec:interview}), complemented by an extensive online study (Sec. \ref{sec:online}), involving a representative subset of 20 of our collected real-world GIFs in order to reduce the study's complexity. 
The studies} collect a variety of data including observation, think-aloud protocols, questionnaires, and subjective feedback. 
The results indicate that many design factors have an impact on the understandability of data-GIFs.
In the end, we summarize a set of design suggestions for creating more effective data-GIFs, and discussed the limitations and future research directions. 
\hl{All materials including the entire GIF corpus, labeled design space, and supplementary materials for interviews and online studies can be found online: \url{https://data-gifs.github.io}.} 
The major contributions are:
\begin{compactitem}[$\diamond$]
    \item The structured design space based on a corpus of \num{} data-GIFs, which summarizes current practices and extracts key design factors.
    \item Exploratory user studies that gain insights into the effect of different data-GIF designs from the standpoints of audiences. 
    \item A set of design suggestions for creating more effective data-GIFs.
\end{compactitem}
\section{Related Work}
We situate our work related to research on data-driven storytelling, animated GIFs and visualization, and studies to measure understandability.

\subsection{Genres in Data-driven Storytelling}
Narrative visualization has been widely used to communicate data insights to the public. 
Segel and Heer \cite{Segel2010} first identified seven genres in 2010: magazine style, annotated charts, posters, flow charts, comics, slideshows, and videos. 
Since then, researches appeared to inform the design of each genre (\textit{e.g.}, \cite{Bach2018,Amini2015, Sarikaya2019, Mei2020vi,Tang2019}), as well as the generation methods (\textit{e.g.}, \cite{Kim2017,Wang2018,Kim2019,Amini2017,tang2021}). 
In this work, we position data-GIFs as an emerging genre for data-driven storytelling with increasing popularity, which possesses distinct visual features and deserves discussion. 

For example, data-GIFs usually convey a single short message through animated visualization in automatic repetition and without sound. 
Regarding this, data videos consist of complex narrative structures (\textit{i.e.}, establisher, initial, peak, and release) with accompanying audio narration \cite{Amini2015, cao2020vi, junxiu2020}, thus presenting more information and requiring a higher cost from both creators and audiences. 
After further investigation on the design practices, we found obvious differences of design features and strategies between data videos and data-GIFs (as shown in Sec. \ref{sec:survey}), such as animation mapping, context preservation, and repetition. 
On the other hand, data-GIFs with a sequence of frames can help communicate the dynamic process, compared to static single images \cite{Hoffler2007}. 
Moreover, the prevalence of smartphones promotes a design trend that displays data-GIFs on the phone and small multiples on the desktop \cite{Boyer, Brehmer2019}.
Given the differences and growing popularity, we looked into data-GIFs, providing a detailed analysis on the design space and examining their potential for data-driven storytelling.

\subsection{Animated GIFs and Animated Visualization}
Created in 1987, animated GIFs are becoming ubiquitous online in recent years. 
Specifically, Bakhshi et al. \cite{Bakhshi2016} found that animated GIFs were more engaging than other kinds of media such as pictures and videos on the social media platform, Tumblr.  
They also identified several significant factors that contribute to these engaging GIFs, including the animation, storytelling capabilities, and emotion expression. 
Furthermore, some studies \cite{Jou2014,Chen2017} have trained models to predict perceived emotions of viewers for animated GIFs. 
Despite the engagement, Jiang et al. \cite{Jiang2018} found that viewers may have diverse interpretation of animated GIFs in communication. 

In this work, we examined a subset of animated GIFs, \textit{i.e.}, data-GIFs, which are used to convey data-driven stories and are predominantly visualization-based. 
Groege first elaborated the idea of data-GIFs with a small collection of examples \cite{DataGIFs}. 
These examples present different properties from generic animated GIFs which are mainly derived from video clips or image stacking, since data-GIFs are designed to communicate data insights. 
However, few work follows up to study data-GIFs in depth, leaving an unclear design space.
Our work thus takes the step towards this direction with a wider range of data-GIFs, and investigates the underlying stories and designs.  

In practice, data-GIFs commonly incorporate animation in visualization. 
Despite the controversy of using animation for analysis \cite{Robertson2008, Chevalier2014}, researchers generally agree on the advantage of animation for communication \cite{McKenna2017, shu2020}. 
For example, animated representations are useful to convey transitions in statistical data graphics \cite{Heer2007, KimYH2019} \hl{and communicate uncertainty to the public \cite{Kale2019}}. 
Amini et al. \cite{Amini2018} \hl{studied data clips as building blocks to compose longer data videos, and showed that incorporating animation in data clips can improve understandability. 
Instead, data-GIFs present another practice of using animation for storytelling, which concisely narrate a short but full story per se and automatically repeat the animation.
It is different from existing animated visualization and has not yet been evaluated.
Therefore, in this work, we explore the design practice of data-GIFs and examine their communication effects.}

\subsection{Empirical Research \hl{to Measure Understandability}}
\hl{Researches in Information Visualization put great efforts on measuring data visualization comprehension. 
Specifically, researchers develop a multitude of test questions for static data visualizations and tasks to assess visualization literacy in a multiphase procedure \cite{Lee2017vlat}. 
B{\"{o}}rner et al. \cite{borner2015} collected qualitative interview feedback from the general public to analyze their comprehension with static visualization. 
However, these works relate to users’ understandability of static visualization for analysis tasks \cite{cai2018cgf}. 
Our work studies how people interpret an animated data-GIF as a communicative visualization, \textit{e.g.}, which design helps or hinders the understanding of the intended content.} 

\hl{Studies on communicative visualization assess the effects of different designs on viewers’ comprehension \cite{Kong2019}. 
For instance, Bateman et al.~\cite{Bateman2010} measured interpretation accuracy of embellished and non-embellished charts with a set of tests. 
Amini et al.~\cite{Amini2018} asked questions about the content to compare the comprehensibility of animated charts and pictographs with static versions in data videos. 
Finally, Wang et al.~\cite{Wang2019} designed comprehension questions to assess the understandability of data comics with infographics and text; quantitative and qualitative results showed data comics were generally easier to understand. 
Similarly, we collect qualitative descriptions to analyze the effects of various GIF designs on viewers' understandability, complemented by accuracy of comprehension questions from the online study.}


\section{Data-GIF Design in Practice}
\label{sec:survey}
To gain insights into the roles of data-GIFs in data-driven storytelling, we conduct an empirical study to collect real-world data-GIFs and analyze their design practices. 
Our goal is to explore the design space of data-GIFs and capture the specialities of data-GIF designs that might influence the understandability.

\subsection{Survey on Data-GIFs}
Our survey was motivated by the small pilot corpus from Lena Groeger \cite{DataGIFs}, which presented 18 data-GIF examples and classified them according to the content, \textit{e.g.}, \textit{showing the temporal process}, \textit{distribution}, \textit{different views}, or \textit{little stories}.  By inspecting this collection, we formulated an initial understanding of data-GIF designs and considered their roles in data storytelling. 
Therefore, we expanded the corpus with 40 data-GIFs published by the leading media outlets on the social media platforms, news websites, and blogs, such as \textit{The New York Times}, \textit{Financial Times}, \textit{Flowing Data}, and \textit{Google Trends}. 
To further inform the design space, we collected another 50 data-GIFs from heterogeneous sources through Google advanced image search for GIFs with keywords including ``\textit{data}'', ``\textit{visualization}'', ``\textit{statistics}'', and ``\textit{graphics}''. 

We established three selection criteria to improve the representativeness of our corpus.
First, it should convey insights supported by data and contain at least one data visualization.
Second, as our motivation pertains to studying data-GIFs as a distinct narrative visualization genre, we excluded GIFs for system demonstrations and animated infographics \cite{Jacob}. 
These GIFs do not leverage this format for the purpose of data-driven storytelling. 
Third, we did not collect duplicate or templated forms of data-GIFs in the course of the survey, since the corpus aims to span a wide range of visualizations types, animation designs, and data stories in data-GIF practices. 

As a result, we arrived at a corpus of 108 data-GIFs.
While not necessarily fully representative, our corpus presents necessary empirical evidence to analyze the designs in current practices.
The complete corpus can be found on the website along with the original sources and labeled design factors (as described in Sec. 3.3 and 3.4).

\subsection{Data Analysis}
\begin{figure*}[htb]
\setlength{\belowcaptionskip}{-7pt} 
\center
    \includegraphics[width=\linewidth]{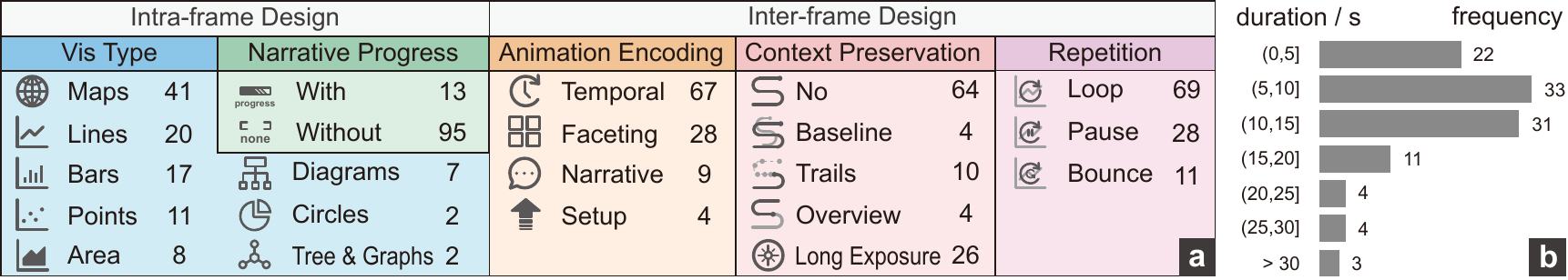}
    \caption{An overview of the design space (a) and the duration distribution (b) of 108 data-GIFs in our survey.}~\label{fig:overview}
\end{figure*}

Informed by our survey, we aimed to explore the design practices and capture designs factors that might contribute to understandability.
We first computed the playback duration, which is a metadata feature (included in the file by default) of animated GIFs and proven to have a strong impact on the interpretation and engagement \cite{Bakhshi2016,Jiang2018}. 
Overall, the average duration of the collected data-GIFs is 11.87 seconds (ranging from 1.4 seconds to 63 seconds), while 78.7\% of the total data-GIFs (85/108) last less than 15 seconds (suggested as the upper limit for animated GIFs \cite{duration}), \hl{as shown in Fig.~\ref{fig:overview}b}. 

Furthermore, we conducted a qualitative analysis to study the content-based design factors of a data-GIF from two aspects: 
\begin{compactitem}[$\bullet$]
    \item \textbf{Intra-frame design:} \textit{What visuals do data-GIFs commonly incorporate in each static frame to present content?} and
    \item \textbf{Inter-frame \hl{design}:} \textit{How do designers articulate the connection between frames and craft animations based on the GIF format? }
\end{compactitem}

Subsequently, we drilled down into each aspect and captured specific design factors.
The whole design space was developed through an iterative process, which started with an analysis of the 18 examples \cite{DataGIFs}, improved with several rounds of discussion among the authors for the growing corpus, and finally refined with user feedback. 
Three of the authors went through all the collected data-GIFs and completed the coding based on the predefined scheme individually, whereby mismatches were resolved through discussions. 
Finally, we revised the coding according to the feedback from the user studies.  
We introduce our resulting five design factors (F1-F5) in the following two sections.

\begin{figure*}[hb]
    \includegraphics[width=\linewidth]{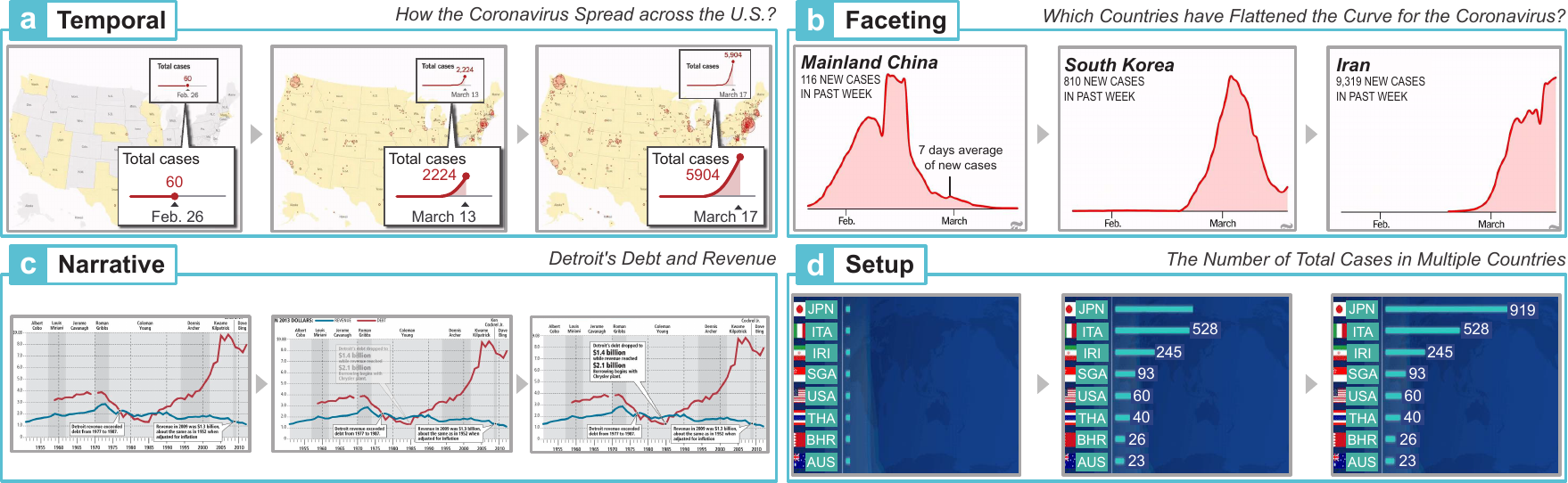}
    \caption{Four data-GIFs with tailored keyframes demonstrating four animation types of data-GIFs. Text is enlarged to make the figure clear. (a) \textbf{Temporal}: It shows the evolvement of the coronavirus in U.S. over time \cite{temporal}. Shadowed callouts are added by paper authors to make the line chart clear. (b) \textbf{Faceting}: It presents the curves for the coronavirus of several countries one by one \cite{faceting}. (c) \textbf{Narrative}: It narrates a story by building up the visualization scene \cite{narrative}. (d) \textbf{Setup}: It animates the creation of a bar chart \cite{setup}. Original GIFs are attached in the supplementary materials. }~\label{fig:animation}
\end{figure*}

\subsection{Intra-frame Design}
Traditionally, GIFs are a consecutive sequence of frames played in loops.
Researches on animated GIFs commonly analyze their content with regard to each frame. 
For example, prior studies \cite{Bakhshi2016, Jou2014} compute the per-frame content features of animated GIFs, such as the face numbers and regions, to examine their impacts on communication and engagement. 
In the context of data-driven storytelling, we consider the content features for each frame from the perspectives of visualization types and narrative progress. 

\begin{figure*}
    \setlength{\belowcaptionskip}{-7pt}
    \includegraphics[width=\linewidth]{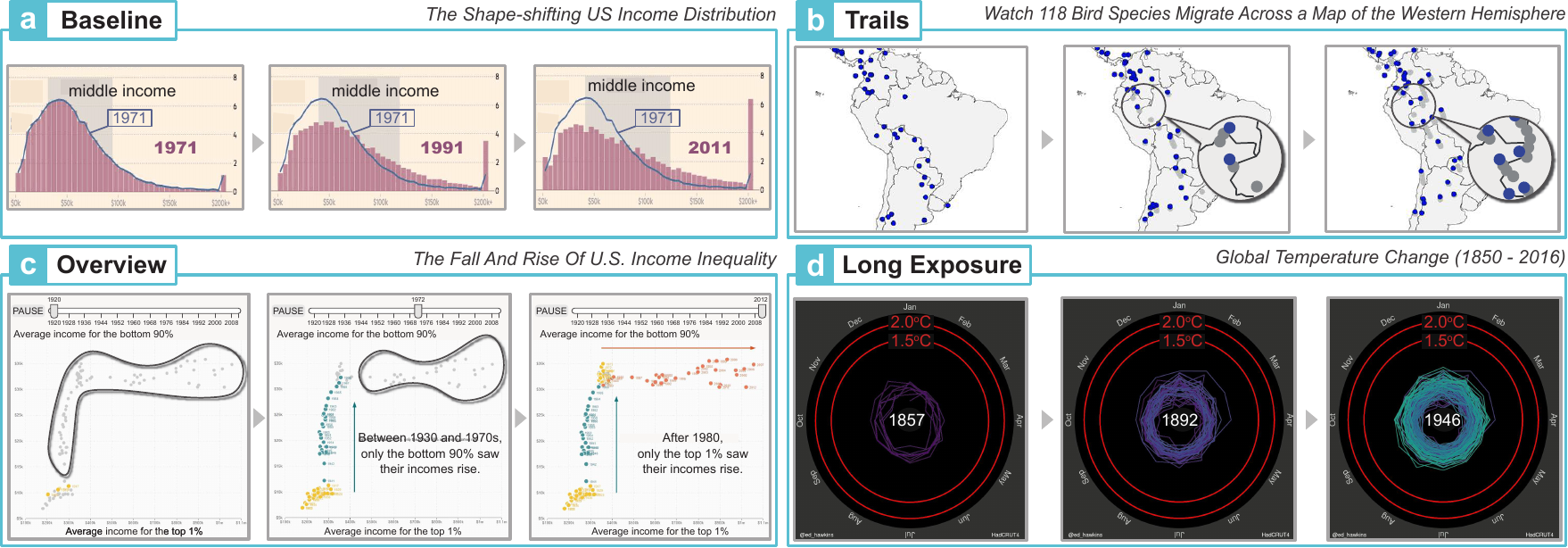}
    \caption{Four data-GIFs with tailored keyframes demonstrating different context preservation techniques of data-GIFs. Shadowed circles in (b) and (c) are our annotations. (a) \textbf{Baseline}: It preserves the distribution of the first year 1971 with the blue line \cite{baseline}. (b) \textbf{Trails}: Each point for a bird leaves the gray trails to show the migration trajectory \cite{trails}. (c) \textbf{Overview}: The gray points show an overview of the data, where each point indicates the value for an upcoming year \cite{overview}. (d) \textbf{Long Exposure}: The colored line for the temperature change is growing spirally, and overlays on the previous \cite{long}. We provide the original GIFs in the supplementary materials. }~\label{fig:context}
\end{figure*}

\subsubsection*{\textbf{F1: Visualization Types}}
Visualizations play a primary role in data-GIFs for conveying insights with data.
We first gain an overview of the visualization usage in a GIF and calculate the number of different visualizations for each data-GIF. 
We find that most data-GIFs (86.1\%; 93) consist of exactly one type of charts, and the rest contain two types.  
This finding is much different from those obtained for data videos, which usually include multiple scenes with different visualizations \cite{Amini2015}.
It suggests that data-GIFs tend to have lower information density than data videos because of the limitation of playback duration and communication modes.

Furthermore, we classify the visualization types based on Borkin et al.'s taxonomy \cite{Borkin2013}.
Visualization types are relatively limited in data-GIFs, where basic charts predominate the corpus. 
As shown in Fig. \ref{fig:overview}, maps (38.0\%; 41) were the most commonly used by a large margin, a difference in frequency compared to static visualizations.
Line (18.5\%; 20) and bar (15.7\%; 17) charts follow. 
However, pictograms which are common in data videos \cite{Amini2015} have rarely appeared in data-GIFs (4/108). 
In summary, data-GIFs embrace simple and intuitive visualization designs. 
This might be motivated by the needs to reduce cognitive load, since GIFs could include overwhelming information but do not allow users to control the pace and pause.

\subsubsection*{\textbf{F2: Narrative Progress}}
\label{sec:progress}
In addition to visualizations, some data-GIFs incorporate designs of narrative progress within each frame, which describes how viewers identify the current playback progress (similar to navigation progress in visual narrative flows \cite{McKenna2017}).
For example, as shown in Fig. \ref{fig:animation}a, the GIF provides a supplemental line chart which not only shows the increase of the total cases but also indicates the progress over time. 
Another example (Fig. \ref{fig:repetition}c) directly uses a timeline to showcase the progress. 
However, only a few data-GIFs (13/108) have integrated such obvious navigation progress designs, thus requiring viewers to perceive the playback progress themselves. 
This could be difficult especially under the scenario of automatic repetition.  

\subsection{Inter-frame \hl{Design}}
This perspective captures the animated features of data-GIFs for articulating the connection between frames. 
Designers leverage the natural properties of the GIF format, such as the temporal context and automatic loops, to improve the elaboration of data stories. 
Considering the designs to describe the content relations within a repetition or between repetitions, we have identified the following three factors. 

\subsubsection*{\textbf{F3: Animation \hl{Encoding}}}
Frames in the data-GIF are arranged in a specific sequence to convey stories, leading to the question, \textit{``what does the data-GIF show when the GIF is playing?''}
We examine this by considering the information encoded by the GIF playing progress. 
We find that the animation can be divided into two categories, \textit{i.e.}, for temporal (62.0\%; 67) and non-temporal meaning (38.0\%; 41).
This might be due to the wide utilization of animation for tracking changes over time \cite{Hoffler2007, Chevalier2016}. 
Therefore, most data-GIFs in our corpus are found to convey a temporal process. 
Referring to content relation in other narrative visualization genres \cite{Bach2018,Amini2018}, we further extract three different types of animations in non-temporal data-GIFs, namely, faceting, narrative, and setup. 

\begin{itemize}[itemsep=0pt,partopsep=0pt,parsep=\parskip,topsep=1pt,leftmargin=*]
    \item \textbf{Temporal} (62.0\%; 67) --- The majority of data-GIFs communicate temporal changes of a data set. One possible explanation might be that people naturally link the temporal context with the GIF playing progress, thereby resulting in a large number of data-GIFs in this category. Specifically, some GIFs (Fig. \ref{fig:animation}a) describe the development of data in a continuous, chronological sequence, and others (Fig. \ref{fig:context}a) may present data in multiple specific moments or time periods.  

    \item \textbf{Faceting} (26.0\%; 28) --- Another portion of data-GIFs are designed to show the facets of a data set in a series of frames respectively (refer to the faceting pattern in data comics \cite{Bach2018}). For example, different data items are encoded in a set of frames regarding the same attribute for comparison. Fig. \ref{fig:animation}b shows the curves of different countries for the coronavirus successively. In addition, other GIFs can present different attributes of a data item that deliver complementary views. 
    
    \item \textbf{Narrative} (8.3\%; 9) --- This type builds a narration during the animation~\cite{Bach2018}, where it introduces problems, provides data contexts, and complements explanatory texts. Most are revealed in a step-by-step presentation, guiding viewers' attention along with the narrative through highlighting and annotating as shown in Fig. \ref{fig:animation}c. 
    
    \item \textbf{Setup} (3.7\%; 4) --- Data-GIFs in this portion animate the creation of a visualization. 
    The term learns from the setup animation in data videos \cite{Amini2018}. 
    They only build the visualization scene but do not encode data by the animation.
    For example, Fig. \ref{fig:animation}d shows that the bars are growing and finally presents the number of total cases in multiple countries. 
    The process and speed do not encode information. 
\end{itemize}


\subsubsection*{\textbf{F4: Context Preservation}}
While animation is commonly criticized for not allowing viewers to track the process, we find there exist designs in data-GIFs that help reveal the connection among frames and keep the reading progress contextualized. 
In this aspect, we summarize the techniques used in our corpus of data-GIFs to answer the question, \textit{``how can viewers track the previous data within a loop?''}
We describe the techniques with regard to the extents of context preservation, i.e., \textit{no} context preserved (59.2\%; 64), \textit{partial} context preserved (16.7\%; 18), and \textit{entire} context preserved (24.1\%; 26). 
Examining into the corpus, we further capture three different techniques in partial context preservation, namely, \textit{baseline}, \textit{trails}, and \textit{overview}.

\begin{figure*}
    \setlength{\belowcaptionskip}{-7pt}
    \includegraphics[width=\linewidth]{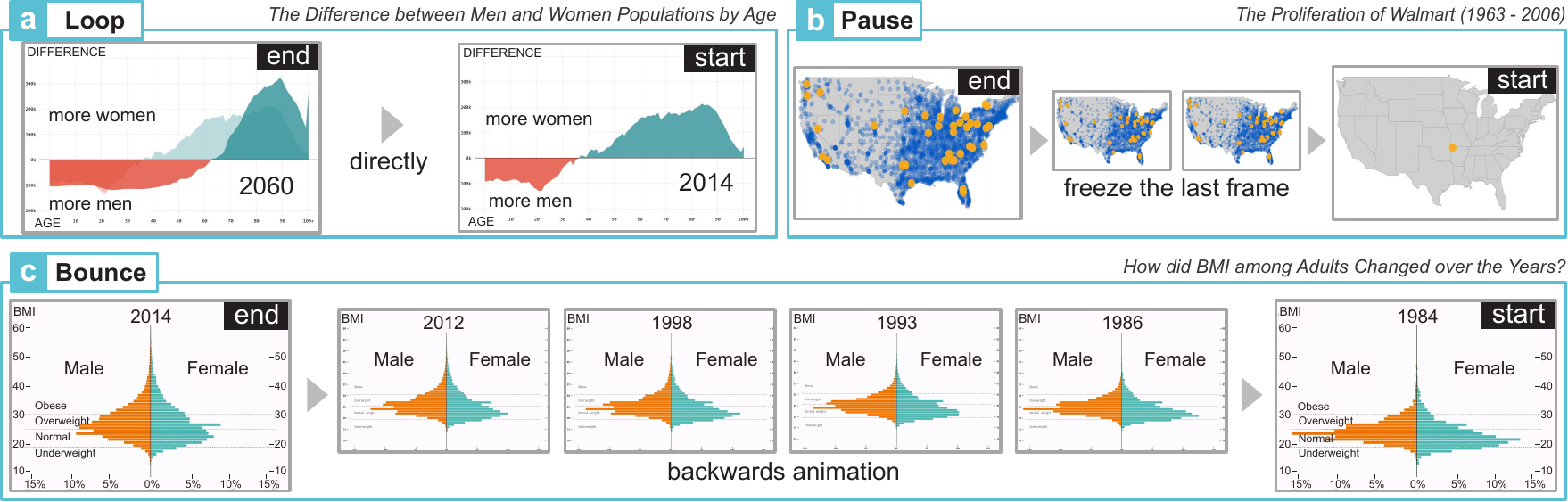}
    \caption{Three data-GIFs with tailored keyframes demonstrating three repetition techniques of data-GIFs. Some text is enlarged to make the figure clear. (a) \textbf{Loop}: It directly starts a new loop \cite{loop}. (b) \textbf{Pause}: It inserts several same frames at the end before starting the next loop \cite{pause}. (c) \textbf{Bounce}: It inserts several frames at the end of the loop, showing backwards animation \cite{bounce}. We provide the original GIFs in the supplementary materials.}~\label{fig:repetition}
\end{figure*}

\begin{itemize}[itemsep=0pt,partopsep=0pt,parsep=\parskip,topsep=1pt, leftmargin=*]
    \item \textbf{No context preservation} (59.2\%; 64) --- Most data-GIFs just play the animation straight through and do not preserve previous data, thus requiring the mental memory of viewers to follow the GIF. For example, Fig. \ref{fig:animation}b switches among different countries and only presents the current country per frame. 
    \item \textbf{Baseline} (3.7\%; 4) --- Baseline designs freeze the content in the first frame of the loop as a baseline during the animation, thereby allowing for comparison with the later frames. Typically, it can directly preserve the first frame and adjust the opacity to alleviate clutter, as shown in Fig. \ref{fig:repetition}a. Others may change the representation. For example, Fig. \ref{fig:context}a replaces the bar chart of the first year with a blue line. 
    \item \textbf{Trails} (9.3\%; 10) --- Trail designs track the data changes between consecutive frames with the GIF playing. In Fig. \ref{fig:context}b, each point leaves a gray trail when it is moving, which indicates its previous positions and speed. Another similar design is the superimposed “trail” variant of Rosling's animation \cite{Robertson2008, Brehmer2019}.
    \item \textbf{Overview} (3.7\%; 4) --- Several examples incorporate an overview in the frame sequence, which presents a summary of data. For example, in Fig. \ref{fig:context}c, the gray points shown from the beginning foreshow the upcoming points and guide the anticipation of the viewers. 
    \item \textbf{Long exposure} (24.1\%; 26) --- This naming is borrowed from the term in photography, which retains information from previous frames. In data-GIFs, long exposure overlays data from all previous frames on the current frame with the GIF playing (\textit{i.e.}, \textit{entirely} context preservation). As shown in Fig. \ref{fig:context}d, the colored line is growing spirally, and the new growth overlays on the previous one. 
\end{itemize}

\subsubsection*{\textbf{F5: Repetition}}
GIFs are particularly featured with automatic repetition, which can influence the reading experience of viewers~\cite{Bakhshi2016}.
We consider the between-loop design and examine the problem, \textit{``what happens after the GIF is played once?''}
We identified three different approaches to end the loop and start a new one. 
\begin{itemize}[itemsep=0pt,partopsep=0pt,parsep=\parskip,topsep=1pt, leftmargin=*]
    \item \textbf{Loop} (63.9\%; 69) --- Most data-GIFs do not have obvious designs that enable viewers to identify the end or the start of a certain loop. Specifically, some examples (Fig. \ref{fig:repetition}a) directly start a new loop once the last loop ends, while several GIFs transition from the end of the last loop to the start of the new loop, completing a seamless transition between two consecutive loops \cite{faceting}.  
    \item \textbf{Pause} (25.9\%; 28) --- Some data-GIFs deliberately pause a while before starting the new loop, thus forming a ``freeze'' of the last frame. This technique helps viewers to clearly identify each repetition. Fig. \ref{fig:repetition}b shows that several frames are inserted at the end of the last loop. 
    \item \textbf{Bounce} (10.2\%; 11) --- A particular design is ``bouncing'', where data-GIFs play the animation once and then reverse it to the start state of the loop. It works similarly to tracing back the history. For example, Fig. \ref{fig:repetition}c designs several frames with backwards animation between the end and the start of the loop. 
\end{itemize}

\section{Semi-structured Interviews}
\label{sec:interview}
The above analysis presents a variety of design choices in visuals and animation, leading to the question, \textit{``What makes a Data-GIF understandable to its intended audience?''} 
Therefore, we conducted our first user study through a series of semi-structured interviews, which aimed to investigate \textit{G1)} how viewers read and understood data-GIFs in the nature of automatic repetition, and \textit{G2)} how each factor influences comprehension.  
We recorded viewers' reading experience in a think-aloud approach and analyzed their qualitative feedback. 


\subsection{Stimuli}
\label{sec:stimuli}
\hl{In this study, we aimed to understand viewers' comprehension of different data-GIFs and their design decisions. To that end, we selected GIFs representing a diverse range of designs from our design space (Sec.~3).
By referring to similar methodologies \cite{Bach2016, Peck2019, Borkin2013} and through several rounds of small-scale pilot studies, we decided on a subset of 20 data-GIFs (approximately 20\% of the total data-GIFs in our corpus) as stimuli in our user studies (see supplementary materials. 11 of them are shown in Figures \ref{fig:animation}, \ref{fig:context}, and \ref{fig:repetition}). 
The frequency of each design in our 20 sample GIFs is shown in Fig. \ref{fig:results}. }

\hl{We acknowledged that in choosing representative GIFs, not all designs could be accommodated to an equal number, making the samples less a less controlled set with respect to the designs. However, the major considerations were threefold and our choice reflects a fair distribution of designs across our entire collection. First, we wanted to test real-world GIFs and the distribution of designs in practice is unbalanced and there were only several example GIFs in specific categories. 
Second, the construction of a data-GIF is complicated, and altering a design factor can inevitably influence others. For example, it is hard to change a \textit{faceting} data-GIF to show a temporal process while preserving other factors. It is hard to untangle all the characteristics and propose fully controlled samples. Third, to our knowledge, no prior study on data-GIFs indicate which design factors are more worth studying and comparing. 
We consider our study as a first step towards the understanding of data-GIFs, and collect general observations and preliminary statistics. Future studies can use our design space to manually  generate data-GIFs and control for individual designs.}


\subsection{Participants and Experimental Setup}

\textbf{Participants.} We recruited 18 participants by disseminating advertisements through emails and at online social groups. 
To ensure that the participants have experience in animated GIFs and have a certain level of visualization literacy, they were pre-screened through several self-reported questions in emails before finally inviting them to the interviews. 
As a result, we recruited 18 participants (8 males and 10 females, aged ranging from 19 to 28, mean: 23.7). 
They reported their frequency in reading or using animated GIFs on social media platforms (6 daily, 9 every three or four days, and 3 weekly). 
The familiarity with data charts was diverse (\textbf{1} \score{} \textbf{7}), but all had a basic visualization literacy. 
Participants came from various backgrounds, including visualization postgraduates (4), digital media students (3), information engineering students (3), environment science students (2), software engineers (2), financial practitioners (2), UX designer (1), and government servant (1).
The participants were represented as P1-P18 in the paper, respectively. 
They were rewarded \$10 compensation for the interview, independent of their performance.

\textbf{Procedure.} 
The study included two major parts, \textit{i.e.}, a GIF reading and describing part, and a follow-up interview part. 
At first, we briefly introduced data-GIFs and \hl{provided a training session, in which we went through the procedure with one data-GIF example}. 
Then, we asked each participant to read and describe 10 data-GIFs in total, since we intended to control the interview duration within 1 hour and keep participants active and engaged.  
The data-GIFs for each participant are randomly selected from the above 20 stimulus.
For each GIF, we deliberately banned automatic repetition and instead asked participants whether they wanted another loop. 
The intention was to capture viewers' reading experience in each loop. 
\hl{We decided on this procedure, based on our pilot study where participants were first allowed to read GIFs freely (\textit{i.e.}, repeat automatically), but we found that they did not say much in the first several loops and began to talk when they had an initial idea, which was not as expected.} 
Specifically, we first played the GIF once, then hid the GIF and asked participants to describe as much as what they had seen, \textit{e.g.}, visual variables and their meanings.
During the description, they were encouraged to verbalize any insights they saw on the GIF, as well as their confusions and comments towards the design. 
\hl{After their description, we would ask whether they wanted another loop. 
If they said like \textit{``I want to see it one more time''},} we played the GIF once again and repeated the description task. 
Otherwise, they might reply \textit{``I cannot find anything more''} or \textit{``I think I've got all the information''} and move to the next GIF.
After reading 10 GIFs, we asked participants about their opinions on data-GIFs, \textit{e.g.}, what contributes to their understanding and which roles data-GIFs play compared with other media. 
All the interviews were audio-recorded and lasted 50 minutes on average. 
The study was run on the Chrome browser at a 13-inch MacBook Pro.

\subsection{Findings}
\label{sec:interview-results}
We finally collected 180 responses for total 20 GIFs (9 responses for each GIF) and then decoded their description. 
Informed by the coding scheme for data interpretation talk \cite{Ma2020}, we coded the recordings according to \textit{a)} which visual design they described, \textit{b)} how they interpreted the encoding, and \textit{c)} whether they correctly interpreted it. 
Some interpretations for visual designs were accompanied with another information, \textit{i.e.}, their feedback toward this design. 
It should be noted that we coded viewers' description with regard to each visual encoding, instead of the number of their findings or analytical results. 
This is because data-GIFs can present a core message with multiple additional information, and we did not require participants to find all possible insights. 
We considered the understandability of data-GIFs based on whether the visual encodings could be correctly understood. 
In addition, we corresponded their description to each repetition and counted the total times of repetition for each GIF. 

\subsubsection*{Goal 1: Reading Experience}
\label{sec:experience}
To gain insights into viewers' reading experience of data-GIFs, we began with analyzing repetition times and comparing differences of viewers' descriptions between each repetition. 
134 responses (74\%) played the GIF more than once, with 77 responses (31\%) played exactly twice. 
The average repetition for 20 GIFs was 2.17 times. 
It showed that viewers were likely to engage in the repetition (P7:\textit{``One more time. I wanted to see whether there are any other interesting things''} and P3:\textit{``I would like to check my understanding''}). 
In the follow-up interviews with those asking for another repetition, when we asked whether they would replay the content if it was a video, all of them reported no (P13: \textit{``It was troublesome especially when it was on the mobile''} and P7: \textit{``Since it was automatically repeated that I cannot help but read it one more time''}). 
P2 further explained \textit{``but if it was important information, I would replay the video. The reason may be that the information the GIF currently shows is not essential to me''}. 

However, this behavior only existed in those easy-to-understand data-GIFs which could be interpreted by viewers within the first or second loops and formed an initial idea about the conclusion. 
Among those responses with more than 3 repetitions (8\%), most struggled to understand the GIF and thus felt upset and bored, let alone further exploration (P8: \textit{``It was terrible. I still didn't understand even after five loops.''}). 
We also asked them about their opinions to change the GIF into a video. 
Their responses depended (P13: \textit{``I was not sure. Maybe I would pause to check the content, or I would even not open that.''} and P15: \textit{``If it was a video, I expect an accompanying audio-explanation, otherwise it was same to me.''}). 
In other words, effective data-GIFs could ignite viewers' passion to understand and explore more with additional repetitions, thereby encouraging us to explore the designs that make data-GIFs understandable. 

Next, we compared the differences of viewers' description in each loop. 
Typically, most responses could describe the first correct interpretation at the first repetition. 
Viewers would shift their attention and notice details which were different from their previous focuses in subsequent repetitions. 
By analyzing those responses who did not fully interpret the GIF at the first loop, we found that participants might first guess the meaning of animation based on their experience, and validate their descriptions later by reading the text, but some even misunderstood ultimately.  
They would give a description after each loop, but added like \textit{``I will check the x-axis (or y-axis, text) next to see if it's correct.''}
In addition, a special example was Fig. \ref{fig:repetition}c where both P2 and P17 at first described it showed the relation between ages and weights even if there was no any hints indicating ages. 
Although they corrected their descriptions at the second repetition after reading the labels, we asked about their initial misunderstanding. 
Both said they had read news/stories about the relations between ages and weights before. 
For Fig. \ref{fig:animation}d, P4 still thought it revealed the temporal process of the increase of the cases after watching two repetitions.
When telling her the true meaning, she said she thought it was similar to the bar chart racing. 
This also inspired us to further investigate design factors that would mislead the understanding of data-GIFs. 


\subsubsection*{Goal 2: Insights into the Design Factors}
\hl{\textbf{Viewers show diverse levels of perception on data-GIFs of different animation encoding.}} 
Specifically, identifying the \textit{temporal} process in data-GIFs was commonly not difficult for participants. 
However, it was more challenging for viewers to read a \textit{faceting} data-GIF, which usually required two or three repetitions for them to extract the relationships between frames. 
P14 commented \textit{``The transition between frames did not help me identify the connections. I had to remember.''}
\textit{Narrative} data-GIFs were considered as step-by-step presentations.  
P1 enjoyed the way the GIF (Fig. \ref{fig:animation}c) \textit{``unfolds an embellished visualization in the narrative timeline.''} 
However, we found \textit{setup} data-GIFs (Fig. \ref{fig:animation}d) can be ambiguous, which \hl{might be mistakenly linked to \textit{time}.} 
P4, P11, and P17 all thought it revealed a temporal process.  
P17 said, \textit{``Cases in Japan grew quickly at first''}, since she related the growing process of the bar to the increasing speed. 
Although P3 correctly interpreted that the GIF was to set up the visualization, she felt disappointed, \textit{``I first thought it would show the temporal process and focused on the speed. But the speed didn't show any meaning, did it?''}
\hl{P2, P3, and P6 all mentioned that it was \textit{``a waste of time''} to play and even repeat the setup animation in GIFs. 
This supports Amini et al.'s suggestion \cite{Amini2018} to use setup animation with care given the delays it can introduce. }

\textbf{Preserving previous data helps identify subtle trends}.
Context preservation designs such as \textit{baseline}, \textit{trails}, and \textit{long exposure} were basically well-received for the ease of comparison. 
This allowed participants to report on detailed trends with these designs. 
For example, in Fig. \ref{fig:context}d, the context (temperature changes of previous years) was preserved through the growing colored line, which made it easy to identify that the temperature did not always increase with a short cooling period. 
P15 further complemented, \textit{``This GIF preserves previous data, so that I could easily find the evolving process.}'' 
However, the preserved context could bring cognitive pressure as well, such as visual clutter and extra encoding interpretation. 
For example, participants felt difficulty in understanding \textit{overview} design (Fig.\ref{fig:context}c) whose gray points indicated the upcoming years, thereby  providing a summary of data. 
P1 said, \textit{``It's useless. I noticed it after I understood the GIF.''}
P3 talked about her confusions at the first repetition, \textit{``what do the color show? the yellow, green, red, and gray points''}.

\textbf{Viewers want the explicit start and end for a repetition}.
Most GIFs do not have a specific designs for repetition, which can make it difficult to differentiate the starting and ending point of a repetition. 
In data-GIFs, many participants did not like this seamless \textit{loop} design (\textit{``I cannot find when the story starts''}). 
In contrast, \textit{pauses} at the end of the loop were appreciated. 
P4 said, \textit{``It gave me time to check the content of the GIF, and I could become accurately aware of a complete loop''}. 
Narrative progress also helped viewers to identify a loop. 
In Fig. \ref{fig:animation}a, the line chart on the top-right corner of the U.S. map shows the increase of the total cases over time. 
P14 said, \textit{``It works as a progress bar and seems to provide an orientation about the progress''}. 
However, for those GIFs (GIF-13 and GIF-14 in the supplementary materials) whose last frame did not subsume important information, some participants thought the \textit{pauses} would not be much useful. 
P15 commented, \textit{``I cannot get any conclusion. The pause (at the last frame) did not provide any help to reflect on the process.''}
Considering the GIF itself, the end frame had low information density, which only indicated a simple final state but lost the information in the progress. 
A possible better design was to incorporate the previous data or take-home messages in the end frame, enabling viewers to gain more information when it paused.   
There also existed the exception which presented the periodic change of population in the Manhattan city. 
The direct \textit{loops} shaped an effect of ``pulse'' and attracted viewers. 
As for \textit{bounce} designs, it could be hard for participants to understand. 
In Fig. \ref{fig:repetition}c, four participants failed in identifying the correct trend after three repetitions (\textit{i.e.}, consider the backwards animation as part of the trend), while two participants at first misunderstood it but corrected later. 
Possible reasons could be that backwards animation was not clearly distinguished and without visual guidance. 
P15 complained, \textit{``why does the GIF need it?''}, while P7 thought it facilitates the comparison between the end and start. 

In summary, we concluded that \textit{1)} participants understood a temporal data-GIF intuitively and might mistakenly link the \textit{setup} animation with the time; \textit{2)} context preservation helped find subtle trends but might exert extra pressure such as visual clutter and context interpretation; \textit{3)} repetition should show the start and end of a loop explicitly. 


\section{Online Study}
\label{sec:online}
Informed by the interviews, we find participants have varied levels of understanding to different data-GIFs, especially for several design factors. 
To examine the differences of different design factors on the understandability, we conduct a large scale online study with representative data-GIFs, aiming to capture specific factors that make a data-GIF understandable and inform the design of effective data-GIFs. 
We use the same set of data-GIFs as those utilized in the interviews for better comparison and analysis of results. 
\hl{The quantitative results should be carefully interpreted given the choice of our 20 GIFs (Sec. \ref{sec:stimuli}).}

\subsection{Hypotheses}
Based on our preliminary user study and literature support \cite{Bakhshi2016,Jiang2017, Borkin2013}, we propose the following hypotheses:
\begin{itemize}[itemsep=0pt,partopsep=0pt,parsep=\parskip,topsep=3pt,leftmargin=*]
    \item \textbf{H-\textsc{Visualization}}: A data-GIF containing basic visualization (\textit{e.g.}, map, bar, and line charts) is most understandable (F1), since viewers could comprehend the visualization easily. 
    \item \textbf{H-\textsc{Progress}:} A data-GIF \textit{with} the narrative progress is more understandable (F2), as it can provide an orientation within the GIF.
    \item \textbf{H-\textsc{Animation}:} A data-GIF showing the \textit{temporal} process is most understandable (F3), since animation are commonly used to convey temporal changes \cite{Heer2007} and viewers intuitively will link animation to time according to our interviewers. 
    \item \textbf{H-\textsc{Context}:} A data-GIF \textit{without} context preservation is least understandable (F4), as it requires viewers' mental map to remember the previous data.  
    \item \textbf{H-\textsc{Repetition}:} A data-GIF with \textit{pause} is most understandable (F5), since it differentiates the start and end of a repetition explicitly according to our interviews. 
\end{itemize}

\subsection{Participants and Experimental Setup}
\label{sec:online-participants}
\textbf{Participants.} 
The experiment was hosted on the Qualtrics survey platform. 
We distributed the survey through multiple methods, \textit{e.g.}, sending emails, inviting participants on visualization-related seminars in a research institution, posting on social media and a campus BBS (Bulletin Board System).
We showcased compelling data-GIFs with a brief introduction and told the duration (approximately 15 minutes) and incentives of the study. 
Participants took part in the study voluntarily, and successful participants were invited to a lottery for a \$5 Amazon voucher.
We provided the reward for the 30\% participants. 
The distribution method and lottery-based incentives were decided to attract more self-directed subjects who participated mainly because of their interests and generate more open-ended feedback \cite{Hsieh2016}. 
In total, 100 participants (mean age: 24.4; SD: 2.77) completed our survey. 
Given participants' ability to read data charts, 4\% indicated no knowledge, 31\% with basic knowledge, 47\% intermediate, and 18\% expert.  
51\% participants reported more than six hours of daily online browsing. 

\textbf{Procedure.} 
The study began with a brief description of data-GIFs, along with the experiment details and its duration (around 15 minutes for total 5 sessions). 
For each session, participants first saw the title of the upcoming data-GIF for 10 seconds, and then watched the GIF (randomly selected from the 20 samples) repeating three times continuously. 
We decided the duration and repetition times based on our experience from previous interviews. 
Later, we hid the GIF, and participants were directed to a questionnaire page. 
\hl{Questions for each GIF included:}
\textit{a)} two 5-point Likert scores for how well they could follow the GIF before and after the questionnaire, ranging from \textit{not at all} (1) to \textit{very well} (5); 
\textit{b)} two open-ended questions listing up elements which helped or hindered the understanding, respectively; 
and \textit{c)} \hl{three multi-choice questions about content and encoding understanding, which were proposed based on the GIFs' source articles and our interview feedback, and pre-tested through pilot studies. 
Each question had five options, including four answer possibilities and \textit{``I don't know''}.
The questions can be found in the supplementary materials.} 
After finishing 5 GIFs, they were asked to fill out a short demographic form. 

\textbf{Data collection.} 
To avoid viewers' prior knowledge influencing the results, we asked whether they have seen any of the GIFs before the questionnaire. 
However, we did not want this question to influence participants' behaviors during the study. 
Thus, we still asked them to complete the questions, but did not consider their responses in our analysis. 
Finally, we had an average of 23.8 responses per GIF (SD: 3.5; total 476 responses for 20 GIFs) \hl{and the average accuracy across all questions and GIFs was 60.1\%}.   
Three types of data from the questionnaire were further analyzed: 
(1) accuracy from multiple-choice comprehension questions; 
(2) subjective scores of viewers' understandings;
and (3) qualitative feedback about influencing design factors.

\begin{figure}
    \setlength{\belowcaptionskip}{-7pt}
    \center
    \includegraphics[width=\columnwidth]{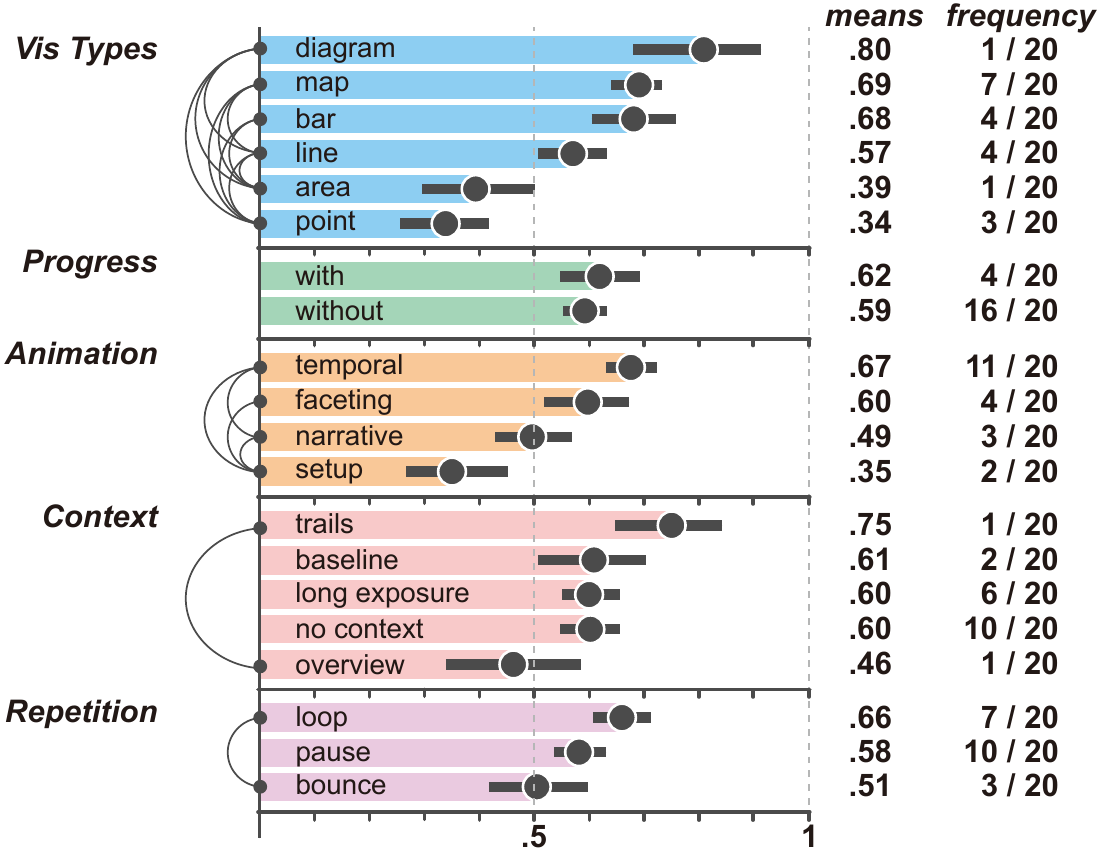}
    \caption{Results for the understandability regarding each design factor with means and 95\% CIs. Frequency shows the appearance times of each design choice in 20 samples. Pair-wise comparisons with statistical significance (after Bonferroni correction) are linked with arcs. }~\label{fig:results}
\end{figure}

\subsection{Findings: What Makes a Data-GIF Understandable?} 
The goal of this analysis is to find the contributing design factors and validate the hypothesis. 
We follow the similar statistical analysis method as Borkin et al. \cite{Borkin2013} and evaluates the statistical significance of the understandability with different design factors. 
In our work, understanding is measured as the accuracy of multi-choice questions in the questionnaire about the GIF content. 
As the accuracy scores in our data are \textit{not} normally distributed, we use \textsc{Kruskal-Wallis} test for each design factor. 
Furthermore, we use \textsc{Wilcoxon two-sided} test for the pair-wise comparison within each factor, and consider Bonferroni correction for multiple comparisons. 
We report the sample means and 95\% confidence intervals in Fig. \ref{fig:results}. 

\paragraph{\textbf{\textsc{Visualization} \textsc{Types}}}
As shown in Fig. \ref{fig:results}, we observe a significant difference between the understanding of data-GIFs and chart types (p\textless0.05). 
\textit{Flow diagrams} \cite{health} show the highest accuracy (mean=.80), although it is not statistically significant compared to maps and bars (p\textgreater0.003). 
After qualitatively viewing participants' responses for elements helping and hindering the understanding, most indicate that both the animation (flowing from the source to the target) and steps (showing each flow one by one) help. 
Possible explanations can be that animation in the flow diagram works as a visual metaphor explaining the encoding of flow diagrams, which facilitates the understanding.  
Meanwhile, many responses also criticize that this GIF design lacks focuses.
On the other hand, we see \textit{area} and \textit{point} charts are significantly less understandable than the others (p-value all less than 0.003). 
Most responses complain that the visualization types are hard to understand (\textit{e.g.}, \textit{``involving multiple variables or dimensions''} and \textit{``cannot understand x (y)-axis''}). 
Back to these GIFs with area (Fig. \ref{fig:repetition}a) and point charts (\textit{e.g.}, Fig. \ref{fig:context}c), we find animation here does not help explain the visualization and even complicates the GIF by adding another data dimension such as time.  
These complex visualization types are less understandable, which we think can help accept H-\textsc{Visualization}. 

\paragraph{\textbf{\textsc{Narrative} \textsc{Progress}}}
Data-GIFs with narrative progress show a slightly higher accuracy (mean=.62) than those without it (mean=.59).
However, there are no significant differences between them (p\textgreater0.05), which rejects H-\textsc{Progress}. 
We check the questions asking about the narrative progress (\textit{``What is the meaning of the line chart?''} for Fig.\ref{fig:animation}a), most responses are correct (81.8\%). 
We also find participants frequently mention the progress bar design in the open-ended questions asking elements which assist in the understanding of those GIFs with narrative progress. 
But some responses criticize that the narrative progress design is \textit{``a little far from the map (the major body of the GIF)''}, which distracts their attention and they have to watch two changing things meanwhile.  
Regarding the roles of narrative progress (\textit{i.e.}, provide the orientation within the GIF), although it is easy to understand and improves the reading experience, viewers still get stuck in understanding other parts of the GIF which are not explained in the narrative progress. 
For example, in Fig. \ref{fig:context}c, the timeline on the top only provides limited help for the understanding of the scatter plots. 

\paragraph{\textbf{\textsc{Animation} \hl{\textsc{Encoding}}}}
Fig.\ref{fig:results} shows a significant difference for animation mapping (p\textless0.001). 
\textit{Setup} is the least accurate with statistical significance (p-value all less than 0.008). 
It is consistent with the interview results that viewers are likely to misunderstand \textit{setup} animation and relate the process to specific data context. 
However, when comparing \textit{temporal} with \textit{faceting}, we could not confidently say \textit{temporal} is better than \textit{faceting} (p=0.053), although the former has a higher mean accuracy. 
To examine the details, we check the scores participants rate for their understanding and find viewers might be less confident about the \textit{faceting} data-GIFs than the \textit{temporal} ones. 
\textit{Faceting} has a lower score (mean=3.12) compared to \textit{temporal} (mean=3.71). 
It corresponds to their qualitative feedback, as many complain it is \textit{``hard to compare between frames''}. 
Consequently, we think \textit{temporal} and \textit{faceting} are the most understandable, \textit{narrative} in between, and \textit{setup} the least understandable, which partially rejects H-\textsc{Animation}.  
Despite no significant difference between \textit{temporal} and \textit{faceting}, viewers are more confident about \textit{temporal} data-GIFs than \textit{faceting}. 


\paragraph{\textbf{\textsc{Context} \textsc{Preservation}}}
We find there is a statistically significant effect of context preservation techniques on the understanding of data-GIFs (p\textless0.05). 
However, through pairwise comparison, we cannot find a context preservation design significantly better than the others (p\textgreater0.005), which also rejects H-\textsc{Context}. 
There are no obvious differences in accuracy for \textit{no context preservation}, \textit{baseline}, and \textit{long exposure} (mean=.60, .60, .61, respectively), while \textit{overview} presents the lowest mean accuracy (mean=.46). 
It indicates that inappropriate context design could be even worse than no context preservation, since it requires attention and efforts for viewers to notice and interpret the preserved context. 
However, the duration of a data-GIF could be short, thus making it hard for viewers to perceive them all at once. 
Especially for \textit{overview}, if viewers could not notice it at first glance, it could not play the expected roles (\textit{i.e.}, presenting data summary), since viewers already see all the data before they realized the \textit{overview} design. 

\paragraph{\textbf{\textsc{Repetition}}}
Results show a significant difference for repetition (p\textless0.05). 
In Fig. \ref{fig:results}, data-GIFs that directly \textit{loop} has a higher accuracy (mean=.65) than those who \textit{pause} (mean=.58, p\textless0.008). 
It is much different from the interview results, and rejects H-\textsc{Repetition}. 
To explain this, we first refer to the comprehension questions posed to participants. 
The questions are designed to \hl{ask} the GIF content related to the repetition, instead of the specific repetition technique (\textit{i.e.}, \textit{pause}, \textit{loop}, or \textit{bounce}). 
It is because explicit questions to identify repetition designs do not contribute to the understanding of the GIF's core message and also influence the viewers' behaviours (\textit{i.e.}, participants who see it once will pay extra attention after that). 
Besides, although we include 10 GIFs with \textit{pauses} and 7 GIFs with \textit{loop} designs to alleviate the bias from other factors, the experiment is still not strictly controlled, which can potentially affect the results. 
When referring to the qualitative feedback from participants, many responses appreciate the \textit{pause} designs can facilitate comprehension and improve the reading experience (\textit{e.g.}, \textit{``pauses give me time to re-think''} and \textit{``it makes me feel better and comfortable''}). 
This reminds us to think about whether \textit{loop} can make viewers more focused on the GIF, thus improving comprehension. 
Moreover, participants' feedback varies given different last frame designs of GIFs. 
For example, they think highly of GIFs that \textit{pause} obviously at the last frame which subsumes important information from the repetition (\textit{e.g.}, the end frame in Fig. \ref{fig:context}d preserves the whole information). 
Thus, viewers could examine what happens from the last frame and re-think the conclusion. 
Otherwise, it \textit{``interrupts my reading and I lose patience when waiting''}, as mentioned in one response. 
Regarding this, we think future studies can conduct a rigorously controlled experiment in the repetition design, since repetition is an essential factor in data-GIFs. 
Better design solutions can be derived from such studies to decide the interval length between two loops and the GIF duration, as well as the design of the last frame. 
\section{Discussion}
This section presents our design suggestions, followed by the discussions about limitations of our study and the roles of data-GIFs. 

\subsection{Design Suggestions}
Based on our findings of the survey and two user studies, we summarize the following design suggestions for creating effective data-GIFs. 
  
\begin{wrapfigure}{l}{0.08\columnwidth}
    \includegraphics[width=0.06\columnwidth, trim=0 0.7em 0.5em 2em]{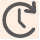}
\end{wrapfigure}
\noindent \textbf{Use animation to convey temporal process.}
Our survey shows that animation in data-GIFs is used to convey different meanings, \textit{e.g.}, \textit{temporal}, \textit{faceting}, \textit{narrative}, and \textit{setup}. 
\hl{Viewers tend to conflate frames with time, even for non-temporal GIFs}, such as regarding \textit{narrative} as a presentation over time.
Meanwhile, despite the similar accuracy between \textit{faceting} and \textit{temporal}, viewers feel more confident to follow a \textit{temporal} data-GIF. 

\begin{wrapfigure}{l}{0.08\columnwidth}
    \includegraphics[width=0.06\columnwidth, trim=0 0.7em 0.5em 2em]{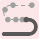}
\end{wrapfigure}
\noindent \textbf{Preserve context.}
Preserving previous data visually supports the comparison and identification of subtle trends. 
However, inappropriate context preservation may cause visual clutter and bring extra cognition pressure that lowers the understandability. 
A suggestion is to carefully consider the context preservation design regarding the data features and intention of the GIF. 
For example, \textit{trails} in Fig. \ref{fig:context}b tally well with the data context (\textit{i.e.}, bird migration), while \textit{long exposure} in Fig. \ref{fig:context}d aims to show the temperature fluctuation. 

\begin{wrapfigure}{l}{0.08\columnwidth}
    \includegraphics[width=0.06\columnwidth, trim=0 0.7em 0.5em 2em]{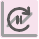}
\end{wrapfigure}
\noindent \textbf{Leverage the end frame \hl{via a pause}.}
In general, we find viewers attach importance to the end frame (\textit{e.g.}, they expect obvious prompts for the end of the repetition and appreciate the \textit{pause} at the end). 
Although the online study shows that the \textit{pause} may not improve the understandability, it does not indicate the uselessness of the end frame.
It instead requires a careful design to make \textit{pause} valuable. 
For example, the \textit{pause} in Fig.\ref{fig:repetition}b allows viewers to examine the previous content.
However, for Fig.\ref{fig:animation}b, even if it \textit{pauses} at the end frame (\textit{i.e.}, showing the curve of a country), it does not bring many benefits for deriving conclusions or recalling contents. 
An idea to leverage the end frame of a GIF is to subsume important information (\textit{e.g.}, previous data or take-home messages) and make it memorable. 

\begin{wrapfigure}{l}{0.08\columnwidth}
    \includegraphics[width=0.06\columnwidth, trim=0 0.7em 0.5em 2em]{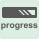}
\end{wrapfigure}
\noindent \textbf{Incorporate narrative progress.}
Integrating the narrative progress design provides the preview and orientation within the GIF.
Although the results do not show a significant effect on the understandability of data-GIFs, viewers express the preference of narrative progress that helps the understanding.
In addition, elaborate narrative progress can also help present the climax (\textit{e.g.}, Fig.\ref{fig:animation}a shows a sharp increase of total cases on the line chart and ignite emotions). 

\begin{wrapfigure}{l}{0.08\columnwidth}
    \includegraphics[width=0.06\columnwidth, trim=0 0.7em 0.5em 2em]{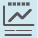}
\end{wrapfigure}
\noindent \textbf{Structure visual content carefully. }
Based on our observations from interviews (Sec. \ref{sec:experience}), viewers' reading experience shares commonality, \textit{i.e.}, \textit{guess-first, details on demand}. 
They will guess the meaning of animation at first glance and formulate an understanding of what they have seen, since animation quickly grabs their attention. 
Then, they will try to examine text and legends. 
If the initial guess confirms, they are likely to begin an active exploration; otherwise, they will struggle. 
Thus, a possible suggestion is to shape data-GIFs in a clear structure that supports a rapid reading and understanding. 

\subsection{Study Limitations and Reflection}
We see our work as a first step toward understanding and assessing (real-world) data-GIFs. 
\hl{The study still has several limitations.} 

\hl{\textbf{Towards controlled studies.}
Our results, especially the quantitative analysis, should be carefully interpreted in the study context of our representative set of 20 GIFs. This selection is decided to reduce the study's complexity and work with real-world GIFs of diverse design factors.
As mentioned in Sec. \ref{sec:stimuli}, we acknowledge that the designs across the evaluated GIFs were limited and may have few samples of a given design, which can affect the results.  
However, it is hard to balance and include all characteristics, especially given the large design space of data-GIFs in both visualization and animation.
Our goal was to run a first informative study using real-world data-GIFs and to inform more controlled studies in future. }

\hl{\textbf{Impact of other design factors.}
In addition to the proposed design factors, other factors may also affect comprehension of data-GIFs.  
For example, visual complexity and the amount of information on comprehension are complicated concepts, especially in animated formats like data-GIFs which present content over time and introduce a multitude of information with the GIF playing. 
Another possible factor might be the duration. 
Our current results suggest that GIFs with relatively long duration are likely to have high accuracy (seen in the supplementary materials). 
However, long GIFs may bring other problems such as engagement, which requires further research.
Finally, showing different GIFs with sometimes similar-but-different design decisions---\textit{e.g.}, the bar grow in \textit{setup} animation is coincidentally similar to the bar chart race animation---may lead to misinterpretations. We acknowledge that this kind of setup animation might not be optimally designed and the results should be taken with care. However, visualization is a field full of such ``false friends''~\cite{wang2020cheat}. 
Other factors like text, information density, and speed of animation may also influence the understandability. 
The exploration on these factors is beyond the scope of this paper, but our study provides many pointers to designing more controlled studies for future works to investigate and quantify these phenomena. 
}

\textbf{Ensuring response quality.}
As mentioned in Sec. \ref{sec:online-participants}, the online study aims to attract self-directed participants who are interested in data-GIFs \cite{Hsieh2016} through multiple methods to improve the response quality.
The descriptive feedback and competition time indicate that most responses are qualified. 
Although the crowdsourcing platform may collect more responses, it is hard to control the experiment context and quality,
\textit{e.g.}, where participants took part in the study as well as other distracting factors \cite{Borgo2018}.
Especially, we cannot directly exclude responses with low accuracy, since misunderstanding responses may indicate ineffective GIF designs in our studies. 
We hope future work can propose more methods to ensure response quality.

\subsection{Data-GIFs for Data-driven Storytelling}

\textbf{Understanding the speciality of data-GIFs.} 
Given the growing popularity of data-GIFs, our initial research question emerges about the fundamental differences between GIFs and other data-driven storytelling mediums.  
From carrying out these explorations in this work, we gain preliminary insights into the particularity of data-GIFs. 

First, data-GIFs are \textit{between pictures and videos}. 
\hl{Based on our interview feedback (Sec. \ref{sec:interview-results}), the GIF format with a sequence of frames and motion effects is considered more suitable to communicate process and attract attention. 
Specifically, most data-GIFs are designed to show a temporal process, especially a spatial-temporal process which can be complicated in one picture. 
Besides, participants appreciated the data-GIFs for \textit{narrative} animation as ``complementing'' a static visualization with step-by-step reading guidance. 
Context preservation techniques track changes among frames, allowing viewers to discern details and extract potential insights from the pacing. 
The characteristics of automatic play and repeat direct viewers' attention to GIFs \cite{Bakhshi2016} and further examine content in multiple loops. 
In addition, the representation of animated GIFs is similar to videos but is shorter and lacks audio-narrated explanation and control of playing progress. These characteristics make GIFs work as a simplified version of videos, which are more accessible with a relative small file size and a single core message.} 
Therefore, we can see data-GIFs as an intermediate format between pictures and videos, where multiple frames enriche content and the video-like animation engages and explains.  

Second, data-GIFs are \textit{between exploration and explanation}. 
\hl{From our interviews (Sec. 4.3), we observed that data-GIFs can encourage viewers to actively explore the details of GIFs in automatic repetition, even after understanding the main message. 
Viewers prefer GIFs that allow them to gain more information from repetitions.
Meanwhile, data-GIFs are a very author-driven medium and are mostly designed to explain authors' messages. }
Thus, data-GIFs work for both exploration and explanation, where the animation drives the viewers to interpret and explore the GIF spontaneously and the content is self-explainable. 

\hl{Despite these, a formative comparison of GIFs to other data-driven storytelling mediums is still required, which would shed light on the opportunities and limitations of GIFs for communication. 
Future works can investigate the design and speciality of data-GIFs deeply.}

\textbf{Data-GIFs as a promising medium.} 
The natural follow-up question is whether data-GIFs can be a promising storytelling medium. 
As animated GIFs are widely used as visual memes in social media and possess characteristics for engagement and virality \cite{Bakhshi2016}, we wonder how data-GIFs leverage the GIF's strengths for attracting and engaging a wide range of audiences. 
Although it is beyond the scope of the paper, we see our findings provide initial evidence. 
For example, viewers responded they would see the GIF repeating and actively explore the details while they may feel reluctant to interact with videos. 
This inspires the research community to further explore the usage scenarios for different storytelling mediums \cite{Bryan2020vi}. 

\textbf{Opportunities for authoring data-GIFs.} 
Another research problem is authoring data-GIFs. 
There are many online GIF maker tools (\textit{e.g.}, GIPHY \cite{duration}), which create animated GIFs by trimming video clips or adding multiple photos. 
However, both approaches are not suitable for creating data-GIFs starting from data.  
The tool support for data-GIFs is still in its infancy. 
Recently, Google published a tool \cite{datagifmaker} that supports the data-GIF creation with only three basic \textit{setup} data-GIFs. 
We hope our corpus and design suggestions could shed light on the future authoring tool design with new and customizable templates. 

\section{Conclusion}
Motivated by the growing popularity of data-GIFs for communicating data stories on social media, we begin by asking what makes a data-GIF understandable to its audiences. 
We conduct a systematic survey of \num{} data-GIFs and two exploratory user studies with total 118 participants that examine the impacts of different design factors on the understanding of data-GIFs. 
Our results identify a set of design factors that have an impact on the understandability of data-GIFs, and also offer valuable design suggestions for creating more effective data-GIFs. 

In this work, we focus on studying data-GIFs in the wild, but further research could generalize the results on larger studies with generated and controlled data-GIFs. 
Beyond understandability, future studies could explore other metrics of data-GIFs such as engagement and socials given the wide use and virality on social media.  
We hope our work could inspire further studies on using and creating data-GIFs for data-driven storytelling, and inform the future authoring tool design. 





\acknowledgments{The authors wish to thank all the data-GIF authors and participants in our studies. The work was supported by National Key R\&D Program of China (2018YFB1004300), NSFC (61761136020), NSFC-Zhejiang Joint Fund for the Integration of Industrialization and Informatization (U1609217), Zhejiang Provincial Natural Science Foundation (LR18F020001) and the 100 Talents Program of Zhejiang University. This project was also supported in part by HKUST SSC grant F0707 and Microsoft Research Asia.}

\bibliographystyle{abbrv-doi}

\newpage
\bibliography{paper}

\begin{thebibliography}{10}

\bibitem{jsvine}
Data {G}{I}{F}s {Collections}.
\newblock Retrieved April 28, 2020 from
  \url{https://www.pinterest.com/jsvine/datagifs/}.

\bibitem{overview}
The {Fall and Rise of U.S. Income Inequality}.
\newblock Retrieved April 28, 2020 from \url{
  https://twitter.com/galka_max/status/823778751079219201}.

\bibitem{flowingdata}
Flowing{Data}.
\newblock Retrieved April 28, 2020 from \url{https://flowingdata.com/}.

\bibitem{nyt}
The {New York Times}.
\newblock Retrieved April 28, 2020 from \url{https://www.nytimes.com/}.

\bibitem{thepudding}
The {Pudding}.
\newblock Retrieved April 28, 2020 from \url{https://pudding.cool/}.

\bibitem{washington}
The {W}ashington {Post}.
\newblock Retrieved April 28, 2020 from \url{https://www.washingtonpost.com/}.

\bibitem{Amini2015}
F.~Amini, N.~Henry~Riche, B.~Lee, C.~Hurter, and P.~Irani.
\newblock Understanding {Data Videos: Looking at Narrative Visualization
  Through the Cinematography Lens}.
\newblock In {\em Proceedings of the ACM Conference on Human Factors in
  Computing Systems}, pp. 1459--1468, 2015. doi: {{%
10\hspace{.1pt}\discretionary{.}{%
}{.}\hspace{.4pt}1145\discretionary{/}{%
}{/}2702123\hspace{.1pt}\discretionary{.}{%
}{.}\hspace{.4pt}2702431}}


\bibitem{Amini2018}
F.~Amini, N.~H. Riche, B.~Lee, J.~Leboe-McGowan, and P.~Irani.
\newblock Hooked on {Data Videos: Assessing the Effect of Animation and
  Pictographs on Viewer Engagement}.
\newblock In {\em Proceedings of the International Conference on Advanced
  Visual Interfaces}, 2018. doi: {{%
10\hspace{.1pt}\discretionary{.}{%
}{.}\hspace{.4pt}1145\discretionary{/}{%
}{/}3206505\hspace{.1pt}\discretionary{.}{%
}{.}\hspace{.4pt}3206552}}


\bibitem{Amini2017}
F.~{Amini}, N.~H. {Riche}, B.~{Lee}, A.~{Monroy-Hernandez}, and P.~{Irani}.
\newblock Authoring {Data}-{Driven Videos} with {D}ata{C}lips.
\newblock {\em IEEE Transactions on Visualization and Computer Graphics},
  23(1):501--510, 2017. doi: {{%
10\hspace{.1pt}\discretionary{.}{%
}{.}\hspace{.4pt}1109\discretionary{/}{%
}{/}TVCG\hspace{.1pt}\discretionary{.}{%
}{.}\hspace{.4pt}2016\hspace{.1pt}\discretionary{.}{%
}{.}\hspace{.4pt}2598647}}


\bibitem{Bach2016}
B.~Bach, N.~Kerracher, K.~W. Hall, S.~Carpendale, J.~Kennedy, and
  N.~Henry~Riche.
\newblock Telling {Stories about Dynamic Networks with Graph Comics}.
\newblock In {\em Proceedings of the ACM Conference on Human Factors in
  Computing Systems}, p. 3670–3682, 2016. doi: {{%
10\hspace{.1pt}\discretionary{.}{%
}{.}\hspace{.4pt}1145\discretionary{/}{%
}{/}2858036\hspace{.1pt}\discretionary{.}{%
}{.}\hspace{.4pt}2858387}}


\bibitem{Bach2018}
B.~Bach, Z.~Wang, M.~Farinella, D.~Murray-Rust, and N.~Henry~Riche.
\newblock Design {Patterns for Data Comics}.
\newblock In {\em Proceedings of the ACM Conference on Human Factors in
  Computing Systems}, pp. 38:1--38:12, 2018. doi: {{%
10\hspace{.1pt}\discretionary{.}{%
}{.}\hspace{.4pt}1145\discretionary{/}{%
}{/}3173574\hspace{.1pt}\discretionary{.}{%
}{.}\hspace{.4pt}3173612}}


\bibitem{Bakhshi2016}
S.~Bakhshi, D.~A. Shamma, L.~Kennedy, Y.~Song, P.~de~Juan, and J.~J. Kaye.
\newblock Fast, {Cheap, and Good: Why Animated GIFs Engage Us}.
\newblock In {\em Proceedings of the ACM Conference on Human Factors in
  Computing Systems}, pp. 575--586, 2016. doi: {{%
10\hspace{.1pt}\discretionary{.}{%
}{.}\hspace{.4pt}1145\discretionary{/}{%
}{/}2858036\hspace{.1pt}\discretionary{.}{%
}{.}\hspace{.4pt}2858532}}


\bibitem{Bateman2010}
S.~Bateman, R.~L. Mandryk, C.~Gutwin, A.~Genest, D.~McDine, and C.~Brooks.
\newblock Useful junk? the effects of visual embellishment on comprehension and
  memorability of charts.
\newblock In {\em Proceedings of the ACM Conference on Human Factors in
  Computing Systems}, p. 2573–2582, 2010. doi: {{%
10\hspace{.1pt}\discretionary{.}{%
}{.}\hspace{.4pt}1145\discretionary{/}{%
}{/}1753326\hspace{.1pt}\discretionary{.}{%
}{.}\hspace{.4pt}1753716}}


\bibitem{narrative}
N.~Bomey and J.~Gallagher.
\newblock Detroit's {Debt}.
\newblock Retrieved April 28, 2020 from
  \url{https://www.freep.com/story/news/local/michigan/detroit/2013/09/15/how-detroit-went-broke-the-answers-may-surprise-you-and/77152028/}.

\bibitem{Borgo2018}
R.~Borgo, L.~Micallef, B.~Bach, F.~McGee, and B.~Lee.
\newblock Information {Visualization Evaluation Using Crowdsourcing}.
\newblock {\em Computer Graphics Forum}, 37(3):573--595, 2018. doi: {{%
10\hspace{.1pt}\discretionary{.}{%
}{.}\hspace{.4pt}1111\discretionary{/}{%
}{/}cgf\hspace{.1pt}\discretionary{.}{%
}{.}\hspace{.4pt}13444}}


\bibitem{Borkin2013}
M.~A. {Borkin}, A.~A. {Vo}, Z.~{Bylinskii}, P.~{Isola}, S.~{Sunkavalli},
  A.~{Oliva}, and H.~{Pfister}.
\newblock What {Makes a Visualization Memorable?}
\newblock {\em IEEE Transactions on Visualization and Computer Graphics},
  19(12):2306--2315, 2013. doi: {{%
10\hspace{.1pt}\discretionary{.}{%
}{.}\hspace{.4pt}1109\discretionary{/}{%
}{/}TVCG\hspace{.1pt}\discretionary{.}{%
}{.}\hspace{.4pt}2013\hspace{.1pt}\discretionary{.}{%
}{.}\hspace{.4pt}234}}


\bibitem{borner2015}
K.~B{\"{o}}rner, A.~Maltese, R.~N. Balliet, and J.~Heimlich.
\newblock Investigating aspects of data visualization literacy using 20
  information visualizations and 273 science museum visitors.
\newblock {\em Information Visualization}, 15(3):198--213, 2016. doi: {{%
10\hspace{.1pt}\discretionary{.}{%
}{.}\hspace{.4pt}1177\discretionary{/}{%
}{/}1473871615594652}}


\bibitem{Boyer}
B.~Boyer.
\newblock Data viz solutions: small multiples on desktop, gifs on yer phone!,
  2015.
\newblock Retrieved April 28, 2020 from
  \url{https://twitter.com/brianboyer/status/583311823245561856}.

\bibitem{Brehmer2019}
M.~{Brehmer}, B.~{Lee}, P.~{Isenberg}, and E.~K. {Choe}.
\newblock A {Comparative Evaluation of Animation and Small Multiples for Trend
  Visualization on Mobile Phones}.
\newblock {\em IEEE Transactions on Visualization and Computer Graphics},
  26(1):364--374, 2020.

\bibitem{Bryan2020vi}
C.~Bryan, A.~Mishra, H.~Shidara, and K.-L. Ma.
\newblock Analyzing gaze behavior for text-embellished narrative visualizations
  under different task scenarios.
\newblock {\em Visual Informatics}, 2020. doi: {{%
10\hspace{.1pt}\discretionary{.}{%
}{.}\hspace{.4pt}1016\discretionary{/}{%
}{/}j\hspace{.1pt}\discretionary{.}{%
}{.}\hspace{.4pt}visinf\hspace{.1pt}\discretionary{.}{%
}{.}\hspace{.4pt}2020\hspace{.1pt}\discretionary{.}{%
}{.}\hspace{.4pt}08\hspace{.1pt}\discretionary{.}{%
}{.}\hspace{.4pt}001}}


\bibitem{cai2018cgf}
X.~Cai, K.~Efstathiou, X.~Xie, Y.~Wu, Y.~Shi, and L.~Yu.
\newblock A {Study of the Effect of Doughnut Chart Parameters on Proportion
  Estimation Accuracy}.
\newblock {\em Computer Graphics Forum}, 37(6):300--312, 2018. doi: {{%
10\hspace{.1pt}\discretionary{.}{%
}{.}\hspace{.4pt}1111\discretionary{/}{%
}{/}cgf\hspace{.1pt}\discretionary{.}{%
}{.}\hspace{.4pt}13325}}


\bibitem{cao2020vi}
R.~Cao, S.~Dey, A.~Cunningham, J.~Walsh, R.~T. Smith, J.~E. Zucco, and B.~H.
  Thomas.
\newblock Examining the use of narrative constructs in data videos.
\newblock {\em Visual Informatics}, 4(1):8 -- 22, 2020. doi: {{%
10\hspace{.1pt}\discretionary{.}{%
}{.}\hspace{.4pt}1016\discretionary{/}{%
}{/}j\hspace{.1pt}\discretionary{.}{%
}{.}\hspace{.4pt}visinf\hspace{.1pt}\discretionary{.}{%
}{.}\hspace{.4pt}2019\hspace{.1pt}\discretionary{.}{%
}{.}\hspace{.4pt}12\hspace{.1pt}\discretionary{.}{%
}{.}\hspace{.4pt}002}}


\bibitem{Chen2017}
W.~{Chen}, O.~O. {Rudovic}, and R.~W. {Picard}.
\newblock G{I}{F}{G}{I}{F}+: {Collecting Emotional Animated {G}{I}{F}s with
  Clustered Multi-task Learning}.
\newblock In {\em Proceedings of the International Conference on Affective
  Computing and Intelligent Interaction}, pp. 510--517, 2017. doi: {{%
10\hspace{.1pt}\discretionary{.}{%
}{.}\hspace{.4pt}1109\discretionary{/}{%
}{/}ACII\hspace{.1pt}\discretionary{.}{%
}{.}\hspace{.4pt}2017\hspace{.1pt}\discretionary{.}{%
}{.}\hspace{.4pt}8273647}}


\bibitem{Chevalier2014}
F.~{Chevalier}, P.~{Dragicevic}, and S.~{Franconeri}.
\newblock The not-so-staggering effect of staggered animated transitions on
  visual tracking.
\newblock {\em IEEE Transactions on Visualization and Computer Graphics},
  20(12):2241--2250, 2014.

\bibitem{Chevalier2016}
F.~Chevalier, N.~H. Riche, C.~Plaisant, A.~Chalbi, and C.~Hurter.
\newblock Animations 25 {Years Later: New Roles and Opportunities}.
\newblock In {\em Proceedings of the International Working Conference on
  Advanced Visual Interfaces}, p. 280–287, 2016. doi: {{%
10\hspace{.1pt}\discretionary{.}{%
}{.}\hspace{.4pt}1145\discretionary{/}{%
}{/}2909132\hspace{.1pt}\discretionary{.}{%
}{.}\hspace{.4pt}2909255}}


\bibitem{trails}
{Cornell Lab}.
\newblock Watch {118 Bird Species Migrate across a Map of the Western
  Hemisphere}.
\newblock Retrieved April 28, 2020 from
  \url{https://www.allaboutbirds.org/news/mesmerizing-migration-watch-118-bird-species-migrate-across-a-map-of-the-western-hemisphere/}.

\bibitem{baseline}
{Financial Times}.
\newblock The {Shape-shifting US Income Distribution}.
\newblock Retrieved April 28, 2020 from
  \url{https://twitter.com/FT/status/674759218545717252}.

\bibitem{bounce}
Flowing{D}ata.
\newblock Americans are {Growing Bigger}.
\newblock Retrieved April 28, 2020 from
  \url{https://twitter.com/flowingdata/status/742741435275874306}.

\bibitem{loop}
Flowing{D}ata.
\newblock The {Difference between Men and Women Population by Age}.
\newblock Retrieved April 28, 2020 from
  \url{https://flowingdata.com/2018/10/17/ask-the-question-visualize-the-answer/}.

\bibitem{duration}
GIPHY.
\newblock G{I}{F} {Creation} {Best} {Practices}.
\newblock Retrieved April 28, 2020 from
  \url{https://support.giphy.com/hc/en-us/articles/360019914771-GIF-Creation-Best-Practices}.

\bibitem{datagifmaker}
Google.
\newblock Google {Data} {G}{I}{F} {Maker}.
\newblock Retrieved April 28, 2020 from
  \url{https://datagifmaker.withgoogle.com/}.

\bibitem{DataGIFs}
L.~Groeger.
\newblock Data {Visualization} {G}{I}{F}s.
\newblock Retrieved April 28, 2020 from \url{http://lenagroeger.com/datagifs/}.

\bibitem{long}
E.~Hawkins.
\newblock Global {Temperature Change (1850-2016)}.
\newblock Retrieved April 28, 2020 from
  \url{https://twitter.com/ed_hawkins/status/729753441459945474?lang=en}.

\bibitem{Heer2007}
J.~{Heer} and G.~{Robertson}.
\newblock Animated {Transitions in Statistical Data Graphics}.
\newblock {\em IEEE Transactions on Visualization and Computer Graphics},
  13(6):1240--1247, 2007. doi: {{%
10\hspace{.1pt}\discretionary{.}{%
}{.}\hspace{.4pt}1109\discretionary{/}{%
}{/}TVCG\hspace{.1pt}\discretionary{.}{%
}{.}\hspace{.4pt}2007\hspace{.1pt}\discretionary{.}{%
}{.}\hspace{.4pt}70539}}


\bibitem{Hoffler2007}
T.~N. Hoffler and D.~Leutner.
\newblock Instructional {Animation versus Static Pictures: A Meta-analysis}.
\newblock {\em Learning and Instruction}, 17(6):722 -- 738, 2007. doi: {{%
10\hspace{.1pt}\discretionary{.}{%
}{.}\hspace{.4pt}1016\discretionary{/}{%
}{/}j\hspace{.1pt}\discretionary{.}{%
}{.}\hspace{.4pt}learninstruc\hspace{.1pt}\discretionary{.}{%
}{.}\hspace{.4pt}2007\hspace{.1pt}\discretionary{.}{%
}{.}\hspace{.4pt}09\hspace{.1pt}\discretionary{.}{%
}{.}\hspace{.4pt}013}}


\bibitem{Hsieh2016}
G.~Hsieh and R.~Kocielnik.
\newblock You {Get Who You Pay for: The Impact of Incentives on Participation
  Bias}.
\newblock In {\em Proceedings of the ACM Conference on Computer-Supported
  Cooperative Work \& Social Computing}, p. 823–835, 2016. doi: {{%
10\hspace{.1pt}\discretionary{.}{%
}{.}\hspace{.4pt}1145\discretionary{/}{%
}{/}2818048\hspace{.1pt}\discretionary{.}{%
}{.}\hspace{.4pt}2819936}}


\bibitem{health}
{Institute for {Health Metrics and Evaluation}}.
\newblock Flows of {Development Assistance for Health}.
\newblock Retrieved April 28, 2020 from
  \url{https://medicalxpress.com/news/2017-04-widely-disparate-health.html}.

\bibitem{Jiang2017}
J.~A. Jiang, J.~R. Brubaker, and C.~Fiesler.
\newblock Understanding diverse interpretations of animated gifs.
\newblock In {\em Proceedings of the ACM Conference Extended Abstracts on Human
  Factors in Computing Systems}, pp. 1726--1732, 2017. doi: {{%
10\hspace{.1pt}\discretionary{.}{%
}{.}\hspace{.4pt}1145\discretionary{/}{%
}{/}3027063\hspace{.1pt}\discretionary{.}{%
}{.}\hspace{.4pt}3053139}}


\bibitem{Jiang2018}
J.~A. Jiang, C.~Fiesler, and J.~R. Brubaker.
\newblock '{The} {Perfect One': Understanding Communication Practices and
  Challenges with Animated} {G}{I}{F}s.
\newblock {\em Proceedings of the ACM on Human-Computer Interaction},
  2(CSCW):80:1--80:20, 2018. doi: {{%
10\hspace{.1pt}\discretionary{.}{%
}{.}\hspace{.4pt}1145\discretionary{/}{%
}{/}3274349}}


\bibitem{pause}
T.~Jones.
\newblock The {Proliferation of Walmart}.
\newblock Retrieved April 28, 2020 from
  \url{https://twitter.com/toddrjones/status/1167112782124179456}.

\bibitem{Jou2014}
B.~Jou, S.~Bhattacharya, and S.-F. Chang.
\newblock Predicting {Viewer Perceived Emotions in Animated GIFs}.
\newblock In {\em Proceedings of the ACM International Conference on
  Multimedia}, pp. 213--216, 2014. doi: {{%
10\hspace{.1pt}\discretionary{.}{%
}{.}\hspace{.4pt}1145\discretionary{/}{%
}{/}2647868\hspace{.1pt}\discretionary{.}{%
}{.}\hspace{.4pt}2656408}}


\bibitem{Kale2019}
A.~{Kale}, F.~{Nguyen}, M.~{Kay}, and J.~{Hullman}.
\newblock Hypothetical outcome plots help untrained observers judge trends in
  ambiguous data.
\newblock {\em IEEE Transactions on Visualization and Computer Graphics},
  25(1):892--902, 2019.

\bibitem{Kim2019}
N.~W. Kim, N.~Henry~Riche, B.~Bach, G.~Xu, M.~Brehmer, K.~Hinckley, M.~Pahud,
  H.~Xia, M.~J. McGuffin, and H.~Pfister.
\newblock Datatoon: {Drawing Dynamic Network Comics With Pen + Touch
  Interaction}.
\newblock In {\em Proceedings of the ACM Conference on Human Factors in
  Computing Systems}, 2019. doi: {{%
10\hspace{.1pt}\discretionary{.}{%
}{.}\hspace{.4pt}1145\discretionary{/}{%
}{/}3290605\hspace{.1pt}\discretionary{.}{%
}{.}\hspace{.4pt}3300335}}


\bibitem{Kim2017}
N.~W. {Kim}, E.~{Schweickart}, Z.~{Liu}, M.~{Dontcheva}, W.~{Li}, J.~{Popovic},
  and H.~{Pfister}.
\newblock Data-{D}riven {Guides: Supporting Expressive Design for Information
  Graphics}.
\newblock {\em IEEE Transactions on Visualization and Computer Graphics},
  23(1):491--500, 2017. doi: {{%
10\hspace{.1pt}\discretionary{.}{%
}{.}\hspace{.4pt}1109\discretionary{/}{%
}{/}TVCG\hspace{.1pt}\discretionary{.}{%
}{.}\hspace{.4pt}2016\hspace{.1pt}\discretionary{.}{%
}{.}\hspace{.4pt}2598620}}


\bibitem{KimYH2019}
Y.~Kim, M.~Correll, and J.~Heer.
\newblock Designing {Animated Transitions to Convey Aggregate Operations}.
\newblock {\em Computer Graphics Forum}, 38(3):541--551, 2019. doi: {{%
10\hspace{.1pt}\discretionary{.}{%
}{.}\hspace{.4pt}1111\discretionary{/}{%
}{/}cgf\hspace{.1pt}\discretionary{.}{%
}{.}\hspace{.4pt}13709}}


\bibitem{Kong2019}
H.-K. Kong, W.~Zhu, Z.~Liu, and K.~Karahalios.
\newblock Understanding visual cues in visualizations accompanied by audio
  narrations.
\newblock In {\em Proceedings of the ACM Conference on Human Factors in
  Computing Systems}, p. 1–13, 2019. doi: {{%
10\hspace{.1pt}\discretionary{.}{%
}{.}\hspace{.4pt}1145\discretionary{/}{%
}{/}3290605\hspace{.1pt}\discretionary{.}{%
}{.}\hspace{.4pt}3300280}}


\bibitem{Lee2017vlat}
S.~{Lee}, S.~{Kim}, and B.~C. {Kwon}.
\newblock Vlat: Development of a visualization literacy assessment test.
\newblock {\em IEEE Transactions on Visualization and Computer Graphics},
  23(1):551--560, 2017.

\bibitem{Ma2020}
J.~{Ma}, K.~{Ma}, and J.~{Frazier}.
\newblock Decoding a {Complex Visualization in a Science Museum – An
  Empirical Study}.
\newblock {\em IEEE Transactions on Visualization and Computer Graphics},
  26(1):472--481, 2020. doi: {{%
10\hspace{.1pt}\discretionary{.}{%
}{.}\hspace{.4pt}1109\discretionary{/}{%
}{/}TVCG\hspace{.1pt}\discretionary{.}{%
}{.}\hspace{.4pt}2019\hspace{.1pt}\discretionary{.}{%
}{.}\hspace{.4pt}2934401}}


\bibitem{McKenna2017}
S.~McKenna, N.~Henry~Riche, B.~Lee, J.~Boy, and M.~Meyer.
\newblock Visual {Narrative Flow: Exploring Factors Shaping Data Visualization
  Story Reading Experiences}.
\newblock {\em Computer Graphics Forum}, 36(3):377–387, 2017. doi: {{%
10\hspace{.1pt}\discretionary{.}{%
}{.}\hspace{.4pt}1111\discretionary{/}{%
}{/}cgf\hspace{.1pt}\discretionary{.}{%
}{.}\hspace{.4pt}13195}}


\bibitem{Mei2020vi}
H.~Mei, H.~Guan, C.~Xin, X.~Wen, and W.~Chen.
\newblock Datav: Data visualization on large high-resolution displays.
\newblock {\em Visual Informatics}, 4(3):12 -- 23, 2020. doi: {{%
10\hspace{.1pt}\discretionary{.}{%
}{.}\hspace{.4pt}1016\discretionary{/}{%
}{/}j\hspace{.1pt}\discretionary{.}{%
}{.}\hspace{.4pt}visinf\hspace{.1pt}\discretionary{.}{%
}{.}\hspace{.4pt}2020\hspace{.1pt}\discretionary{.}{%
}{.}\hspace{.4pt}07\hspace{.1pt}\discretionary{.}{%
}{.}\hspace{.4pt}001}}


\bibitem{munzner2015}
T.~Munzner.
\newblock {\em Visualization {Analysis and Design}}.
\newblock CRC Press, 2015.

\bibitem{Jacob}
J.~O'Neal.
\newblock Animated {Infographics}.
\newblock Retrieved April 28, 2020 from \url{https://jacoboneal.com/cheetah/}.

\bibitem{Peck2019}
E.~M. Peck, S.~E. Ayuso, and O.~El-Etr.
\newblock Data is {Personal: Attitudes and Perceptions of Data Visualization in
  Rural Pennsylvania}.
\newblock In {\em Proceedings of the ACM Conference on Human Factors in
  Computing Systems}, 2019. doi: {{%
10\hspace{.1pt}\discretionary{.}{%
}{.}\hspace{.4pt}1145\discretionary{/}{%
}{/}3290605\hspace{.1pt}\discretionary{.}{%
}{.}\hspace{.4pt}3300474}}


\bibitem{setup}
{People's Daily}.
\newblock The {Number of Total Cases in Multiple Countries}.
\newblock Retrieved April 28, 2020 from
  \url{https://wap.peopleapp.com/article/5210197/5111953}.

\bibitem{Robertson2008}
G.~{Robertson}, R.~{Fernandez}, D.~{Fisher}, B.~{Lee}, and J.~{Stasko}.
\newblock Effectiveness of {Animation in Trend Visualization}.
\newblock {\em IEEE Transactions on Visualization and Computer Graphics},
  14(6):1325--1332, 2008. doi: {{%
10\hspace{.1pt}\discretionary{.}{%
}{.}\hspace{.4pt}1109\discretionary{/}{%
}{/}TVCG\hspace{.1pt}\discretionary{.}{%
}{.}\hspace{.4pt}2008\hspace{.1pt}\discretionary{.}{%
}{.}\hspace{.4pt}125}}


\bibitem{Sarikaya2019}
A.~{Sarikaya}, M.~{Correll}, L.~{Bartram}, M.~{Tory}, and D.~{Fisher}.
\newblock What do we talk about when we talk about dashboards?
\newblock {\em IEEE Transactions on Visualization and Computer Graphics},
  25(1):682--692, 2019.

\bibitem{Segel2010}
E.~{Segel} and J.~{Heer}.
\newblock Narrative {Visualization: Telling Stories with Data}.
\newblock {\em IEEE Transactions on Visualization and Computer Graphics},
  16(6):1139--1148, 2010. doi: {{%
10\hspace{.1pt}\discretionary{.}{%
}{.}\hspace{.4pt}1109\discretionary{/}{%
}{/}TVCG\hspace{.1pt}\discretionary{.}{%
}{.}\hspace{.4pt}2010\hspace{.1pt}\discretionary{.}{%
}{.}\hspace{.4pt}179}}


\bibitem{shu2020}
X.~{Shu}, J.~{Wu}, X.~{Wu}, H.~{Liang}, W.~{Cui}, Y.~{Wu}, and H.~{Qu}.
\newblock Dancingwords: exploring animated word clouds to tell stories.
\newblock {\em Journal of Visualization}, 2020. doi: {{%
10\hspace{.1pt}\discretionary{.}{%
}{.}\hspace{.4pt}1007\discretionary{/}{%
}{/}s12650\discretionary{%
}{-}{-}020\discretionary{%
}{-}{-}00689\discretionary{%
}{-}{-}0}}


\bibitem{junxiu2020}
J.~Tang, L.~Yu, T.~Tang, X.~Shu, L.~Ying, Y.~Zhou, P.~Ren, and Y.~Wu.
\newblock Narrative {Transitions in Data Videos}.
\newblock In {\em Proceedings of the IEEE Visuliazation Conference Short
  Papers}, 2020.

\bibitem{tang2021}
T.~{Tang}, R.~{Li}, X.~{Wu}, S.~{Liu}, J.~{Knittel}, T.~E. S.~{Koch}, L.~{Yu},
  P.~{Ren}, and Y.~{Wu}.
\newblock Plot{T}hread: Creating {Expressive Storyline Visualizations using
  Reinforcement Learning}.
\newblock {\em IEEE Transactions on Visualization and Computer Graphics},
  27(2):To appear, 2021.

\bibitem{Tang2019}
T.~{Tang}, S.~{Rubab}, J.~{Lai}, W.~{Cui}, L.~{Yu}, and Y.~{Wu}.
\newblock i{S}toryline: Effective {C}onvergence to {H}and-drawn {S}torylines.
\newblock {\em IEEE Transactions on Visualization and Computer Graphics},
  25(1):769--778, 2019.

\bibitem{temporal}
{The {New York Times}}.
\newblock How does the {Coronavirus Spread across the U.S.?}
\newblock Retrieved April 28, 2020 from
  \url{https://twitter.com/nytimes/status/1241387140933652481}.

\bibitem{faceting}
{The {New York Times}}.
\newblock Which {Countries have Flattened the Curve for the Coronavirus?}
\newblock Retrieved April 28, 2020 from
  \url{https://twitter.com/nytimes/status/1240790201414438912}.

\bibitem{Tversky2002}
B.~Tversky, J.~Morrison, and M.~Bétrancourt.
\newblock Animation: {Can it facilitate?}
\newblock {\em International Journal of Human-Computer Studies}, 57:247--262,
  10 2002. doi: {{%
10\hspace{.1pt}\discretionary{.}{%
}{.}\hspace{.4pt}1006\discretionary{/}{%
}{/}ijhc\hspace{.1pt}\discretionary{.}{%
}{.}\hspace{.4pt}2002\hspace{.1pt}\discretionary{.}{%
}{.}\hspace{.4pt}1017}}


\bibitem{Wang2018}
Y.~Wang, H.~Zhang, H.~Huang, X.~Chen, Q.~Yin, Z.~Hou, D.~Zhang, Q.~Luo, and
  H.~Qu.
\newblock Info{N}ice: {Easy Creation of Information Graphics}.
\newblock In {\em Proceedings of the ACM Conference on Human Factors in
  Computing Systems}, 2018. doi: {{%
10\hspace{.1pt}\discretionary{.}{%
}{.}\hspace{.4pt}1145\discretionary{/}{%
}{/}3173574\hspace{.1pt}\discretionary{.}{%
}{.}\hspace{.4pt}3173909}}


\bibitem{wang2020cheat}
Z.~Wang, L.~Sundin, D.~Murray-Rust, and B.~Bach.
\newblock Cheat sheets for data visualization techniques.
\newblock In {\em Proceedings of the ACM Conference on Human Factors in
  Computing Systems}, pp. 1--13, 2020.

\bibitem{Wang2019}
Z.~Wang, S.~Wang, M.~Farinella, D.~Murray-Rust, N.~Henry~Riche, and B.~Bach.
\newblock Comparing {Effectiveness and Engagement of Data Comics and
  Infographics}.
\newblock In {\em Proceedings of the ACM Conference on Human Factors in
  Computing Systems}, 2019. doi: {{%
10\hspace{.1pt}\discretionary{.}{%
}{.}\hspace{.4pt}1145\discretionary{/}{%
}{/}3290605\hspace{.1pt}\discretionary{.}{%
}{.}\hspace{.4pt}3300483}}


\end{thebibliography}
\end{document}